\documentclass[usenatbib]{mn2e}
\usepackage{graphicx,color,hyperref,amsmath,amssymb}

\newcommand{\dtb}{T_b}
\newcommand{\Mpci}{\text{ Mpc}^{-1}}
\newcommand{\Mpc}{\text{ Mpc}}
\newcommand{\MHz}{\text{ MHz}}
\newcommand{\GHz}{\text{ GHz}}
\newcommand{\mJy}{\text{ mJy}}
\newcommand{\muJy}{ \text{ } \mu \text{Jy}}
\newcommand{\nJy}{\text{ nJy}}
\newcommand{\Msun}{ M_{\odot}}

\newcommand{\degree}{^{\circ}}
\newcommand{\kperp}{k_\perp}
\newcommand{\kpara}{k_\parallel}

\newcommand{\be}{\begin{equation}}
\newcommand{\ee}{\end{equation}}
\newcommand{\trans}{\mathsf{T}}
\newcommand{\mK}{\text{ mK}}

\newcommand{\meter}{\text{ m}}
\newcommand{\km}{\text{km}}
\def\Let@{\def\\{\notag\math@cr}}

\begin{document}

\title[Detecting the 21 cm Forest in the 21 cm Power Spectrum]{Detecting the 21 cm Forest in the 21 cm Power Spectrum}

\author[Aaron Ewall-Wice, Joshua S. Dillon, Andrei Mesinger, and Jacqueline Hewitt]{Aaron Ewall-Wice $^{1}$ \thanks{E-mail: aaronew@mit.edu},
Joshua S. Dillon $^{1}$,
Andrei Mesinger $^{2}$,
Jacqueline Hewitt $^{1}$
\\
$^{1}$Dept. of Physics and MIT Kavli Institute, Massachusetts Institute of Technology, Cambridge, MA 02139, USA \\
$^{2}$Scuola Normale Superiore, Piazza dei Cavalieri 7, 56126 Pisa, Italy}

\voffset-.6in

\maketitle

\begin{abstract}
 We describe a new technique for constraining the radio loud  population of active galactic nuclei at high redshift by measuring the imprint of  21 cm spectral absorption features (the 21 cm forest) on the 21 cm power spectrum. Using semi-numerical simulations of the intergalactic medium and a semi-empirical source population we show that the 21 cm forest dominates a distinctive region of $k$-space, $k \gtrsim 0.5 \Mpci$. By simulating foregrounds and noise for current and potential radio arrays, we find that a next generation instrument with a collecting area on the order of $\sim 0.1\km^2$ (such as the Hydrogen Epoch of Reionization Array) may separately constrain the X-ray heating history at large spatial scales and radio loud active galactic nuclei of the model we study at small ones. We extrapolate our detectability predictions for a single radio loud active galactic nuclei population to arbitrary source scenarios by analytically relating the 21 cm forest power spectrum to the optical depth power spectrum and an integral over the radio luminosity function. 
\end{abstract}


\section{ INTRODUCTION}

Observations of emission and absorption at 21 cm  from the neutral intergalactic medium (IGM) at high redshift will offer an unprecedented glimpse into the cosmic dark-ages up through the epoch of reionization (EoR), constraining both fundamental cosmological parameters and the properties of the first stars and galaxies \citep[for reviews]{Furlanetto:2006tc,Morales:2010cb,Pritchard:2012wy}. Direct mapping of the 21 cm signal during the EoR is likely a decade or more away, requiring projected instruments such as the Square Kilometer Array (SKA). However, a first generation of experiments attempting to detect the power spectrum are already underway.  These include the Low Frequency ARray \citep[LOFAR]{2013A&A...550A.136Y}, the Murchison Widefield Array \citep[MWA]{2013PASA...30....7T}, the Precision Array for Probing the Epoch of Reionization \citep[PAPER]{Parsons:2013uu}, and the Giant Metre-wave Telescope \citep[GMRT]{Paciga:2013vb}. The MWA, PAPER, and LOFAR have the potential to achieve statistical detections of brightness temperature fluctuations within the next several years \citep{Bowman:2006tg,McQuinn:2006ty,Beardsley:2013wj,Mesinger:2013c}

Most theoretical investigations of observing neutral hydrogen in the
EoR have focused on IGM emission and absorption
against the Cosmic Microwave Background (CMB). It has
also been recognized by \citet{Carilli:2002tb,Furlanetto:2002,Xu:2009tc,Mack:2012tf,Ciardi:2013to} that the 21 cm forest, HI absorption in the spectra of background radio-loud (RL) active galactic nuclei (AGN), can be used to probe the IGM's thermal state.
  
 Studies of the forest have focused on its detection in the frequency spectra of a known RL source to glean information on the thermal properties of the absorbing IGM. The possibility for such a study depends on the existence of high redshift RL sources. As of 2013, the RL source distribution is only well constrained out to $z \sim 4$ (see \citet{DeZotti:2010} for review). Theoretical work suggests that at 100 $\MHz$ hundreds of $S \sim 1 \mJy$ sources with redshifts greater than 10 might exist within one of the $(30 \degree)^2$ fields of view (FoV) offered by existing and upcoming wide field interferometers  \citep{2004ApJ...612..698H,2008MNRAS.388.1335W} (hereafter H04 and W08 respectively). However the discovery of a suitable source at high redshift entails an extensive follow up program to measure  photometric redshifts of  radio selected candidates. 

Should sufficiently RL sources exist, a line of sight (LoS) detection of individual absorption features will require large amounts of integration time on a radio telescope with the collecting area comparable to the Square Kilometer Array (SKA). At reionization redshifts, \citep{Mack:2012tf} find that a $5 \sigma$ detection of an individual absorption feature with a $z \approx 9$ Cygnus A type source\footnote{ flux density at $151 \MHz$ of $S_{151} \approx 20 \mJy$ and spectral index of $\alpha \approx 1.05$ } would require years of integration on an SKA-like instrument. \citet{Ciardi:2013to} find that after 1000 hours of integration only 0.1\% of the LoS in an IGM simulation box contained regions of large enough optical depth to produce absorption features\footnote{ against a $S_{129} \approx 50 \mJy $ source at $z \sim 7$} observable by LOFAR. Hence a detection of the forest with a present day interferometer would require a very rare juxtaposition of an extremely loud RL source with an outlying optical depth feature. Even if this detection were achieved, it is unlikely that significant inferences on the thermal history could be made from only a handful of such observations.

 While detecting individual absorption features presents an enormous challenge, statistical methods have been demonstrated to reduce the necessary integration times. One target for a statistical detection is the increased variance in flux, along the LoS.  It is shown in  \citep{Mack:2012tf} that the integration time required for detecting this variance increase for a Cygnus A source, is only a few weeks with an SKA-like telescope, as apposed to the decades needed for detecting a single feature. \citet{Ciardi:2013to} find that LOFAR could detect the global suppression in the spectrum of a 50 mJy source at $z \sim 12$ with a 1000 hour integration, though they note that a detection by LOFAR is unlikely due to excessive RFI in the FM band ($80 \MHz \lesssim \nu \lesssim 108 \MHz$).

The possibility of a statistical detection of the forest using information from the wide FoV available to the current and upcoming generations of experiments has not yet been investigated. Observing the forest signature in the 21 cm power spectrum would integrate the signal from many high redshift sources within a FoV, reducing the sensitivity requirements of the instrument. Also, a power spectrum detection does not require a priori knowledge of high redshift sources. Hence the technique we describe can put constraints on both the properties of the IGM, such as the heating and reionization history, and the population of high redshift RL sources. It is likely that 21cm forest absorption features could be fruitfully explored using high-order statistical measures as well, but we do not consider those in this paper.

In this proof-of-concept, we begin to explore the characteristics and observability of the forest in the 21 cm power spectrum. We derive analytically the features that the global forest should introduce to the power spectrum and confirm their existence by combining semi-numerical simulations of the IGM, computed with 21cmFAST \citep{Mesinger:2011tb}, with the semi-empirical model of the high redshift population of RL sources from W08.
We find that in all heating scenarios studied, the contribution to the
21 cm fluctuations by the absorption of our RL sources is comparable to or dominates the contribution from the brightness temperature on small spatial scales $(k \gtrsim 0.50 \Mpci)$. To determine the detectability of the forest in the power spectrum, we perform sensitivity calculations for several radio arrays with designs similar to the MWA, including a future array with a collecting area of $\sim 0.1 \km^2$, similar to the planned Hydrogen Epoch of Reionization Array (HERA). In order to give the reader a sense of how the strength of this signal scales across a large range of of radio loud source populations, we extrapolate the expected S/N of the Forest using our analytic expression for the signal strength.

This paper is organized as follows.  In Section \ref{sec:theory} we provide the theoretical background and use a toy model to derive the morphology of the 21 cm forest power spectrum; relating its shape and amplitude to the optical depth power spectrum and the radio luminosity function. In Section \ref{sec:sim} we describe the semi-numerical simulations of the IGM along with the semi-empirical RL source distribution of W08 and how we combine them to simulate the wide field forest.  In Section \ref{sec:results} we discuss our results and identify the separate regions of $k$-space that may be used to independently constrain the thermal history of the IGM and the high redshift RL distribution. In Section \ref{sec:det} we explore the prospects for detecting the forest in  spherically averaged power spectrum measurements considering the sensitivity of current and future radio arrays. In Section \ref{sec:extr} we extrapolate our detectability results across a broad range of source populations and X-ray heating scenarios. 

Throughout this work we assume a flat universe with the cosmological parameters $h = 0.7$, $\Omega_\Lambda = 0.73$, $\Omega_M = 0.27$, $\Omega_b = 0.082$, $\sigma_8 = 0.82$, and $n=0.96$ as determined by the WMAP 7-year release \citep{Komatsu:2011in}. All cosmological distances are in comoving units unless stated otherwise. 
\vskip 1.0 cm

\section{  THEORETICAL BACKGROUND} \label{sec:theory}

In this section we establish our notation and present a basic mathematical description of how forest absorption modifies the 21 cm brightness temperature signal. 

\subsection{Notation}
We adopt the Fourier transform convention
\begin{equation}
\widetilde{ f }({\bf k}) = \int d^3x   e^{-i {\bf k \cdot x}} f( {\bf x} ). 
\end{equation}
In addition, we often refer to cylindrical Fourier coordinates where $\kperp \equiv \sqrt{k_x^2 + k_y^2}$ and $\kpara \equiv |k_z|$. 
The power spectrum of a field $A$ over a comoving volume V is defined as
\begin{equation}\label{eq:ps}
P_A = \frac{1}{V} \langle | \widetilde{ \Delta A } | ^2 \rangle
\end{equation}
and the cross power spectrum between fields $A$ and $B$ over $V$ is given by
\begin{equation}
P_{A,B}= \frac{1}{V} \langle \widetilde{\Delta A} \widetilde{\Delta B}^* \rangle
\end{equation}
where 
\begin{equation}
\Delta A = A-\langle A \rangle
\end{equation}
and $\langle A \rangle $ is defined as the ensemble average of $A$ though in practice it is computed by averaging over some spatial or Fourier volume. In our discussion, we will also be referring to the one dimensional LoS power spectrum (not to be confused with the 1D spherical power spectrum) of a field A along a LoS column of comoving length L. 
\begin{equation}
P_A^{LoS}(k_z) = \frac{1}{L}  \int dz dz' \Delta A(z) \Delta A(z')  e^{i k_z(z-z')}
\end{equation}
Finally, we use $\Delta^2$ to denote the dimensionless power spectrum
\begin{equation}
\Delta^2 (k) \equiv \frac{k^3}{2 \pi^2} P(k)
\end{equation}
\subsection{The Forest's Modification of the Brightness Temperature}
 The forest absorption traces the optical depth of the IGM and will therefore introduce a signal on similar spatial scales as the 21 cm brightness temperature. We now discuss this signal in detail. The optical depth of a high redshift HI cloud is given by \citep{Furlanetto:2006tc}
\begin{equation}\label{eq:tau}
\tau_{21} \approx .0092(1+\delta)(1+z)^{3/2} \frac{x_{HI}}{T_S} \left[ \frac{H(z)/(1+z)}{d v_{\parallel}/dr_{\parallel} } \right].
\end{equation}
$\delta$ is the fractional baryonic over-density, H(z) is the Hubble factor, $dv_\parallel / dr_\parallel$ is the velocity gradient along the LoS (including the Hubble expansion), and $x_{HI}$ is the neutral hydrogen fraction. The numerical factor in front of Equation (\ref{eq:tau}) is computed from fundamental constants and is independent of cosmology.
The spin temperature, $T_s$ is defined by the relative population densities of the two hyperfine energy levels, $n_1$ and $n_0$  \citep{Field:1958}
\begin{equation}
\frac{n_1}{n_0} = 3 \exp \left( {- \frac{h \nu_{21}}{k_B T_s}} \right).
\end{equation}
 Where, $h$ is Plank's constant, $k_B$ is the Boltzmann constant, and $\nu_{21} = 1420.41 \text{MHz}$ is the rest frame frequency of  the hyperfine transition radiation.

 Prior works on 21 cm tomography assume that the sky temperature at $\nu = \nu_{21}/(1+z)$ in the direction of an HI cloud is given by 
\begin{equation}\label{eq:sky_temp_noRL}
T_{sky} = \frac{T_s}{(1+z)} (1-e^{-\tau_{21}}) + \frac{T_{CMB}}{(1+z)}e^{-\tau_{21}} + T_{fg}.
\end{equation}
where $T_{CMB}$ is the comoving temperature of the cosmic microwave background radiation and $T_{fg}$ is the temperature of foreground emission including synchrotron radiation of the Galaxy, resolved point sources,  free-free emission, and radio emission from unresolved point sources below the confusion limit \citep{2002ApJ...564..576D,2008MNRAS.389.1319J,deOliveiraCosta:2008ks,Wang:2006gm}. 

 The first term in Equation (\ref{eq:sky_temp_noRL}) includes both the 21 cm emission and self absorption of the HI cloud, hence it is multiplied by a factor of $(1-e^{-\tau_{21}})$. The second term describes the observed intensity of a background source shining through the cloud so its temperature is attenuated by $e^{-\tau_{21}}$. The third term describes radiation emitted by sources closer than the cloud so its intensity unaffected by $\tau_{21}$.

 21 cm experiments seek to measure the difference between the first two terms of Equation (\ref{eq:sky_temp_noRL}) and $T_{CMB}$. This difference is often referred to as the ``differential brightness temperature" and is given by \citep{Furlanetto:2006tc}
\begin{equation}\label{eq:dtb_no_RL}
\dtb = \frac{(T_s - T_{CMB})}{(1+z)}(1-e^{-\tau_{21}}) \approx \frac{T_s-T_{CMB}}{(1+z)}\tau_{21}.
\end{equation}
 We depart from previous work by considering the effect of radio loud sources behind the HI cloud whose combined observed\footnote{In accordance with much of the literature, we use the observed temperature for $T_{RL}$ and $T_{fg}$, rather than the comoving temperature as we have for $T_s$ and $T_{CMB}$. As a result, there are no factors of $(1+z)$ under $T_{RL}$ or $T_{fg}$.} brightness temperature we denote as $T_{RL}$. Including these background sources, Equation (\ref{eq:sky_temp_noRL}) becomes
\begin{equation}\label{eq:sky_temp}
T_{sky}' = \frac{T_{s}}{(1+z)} ( 1- e^{-\tau_{21}}) + \frac{T_{CMB}}{(1+z)}e^{-\tau_{21}} + T_{RL} e^{-\tau_{21}} +T_{fg} 
\end{equation}

$T_{fg}$ and $T_{RL}$  are expected to have predominantly smooth spectra which reside within a limited region of Fourier space known as the ``wedge" \citep{Datta:2010he,Morales:2012uk,Vedantham:2012}. Smooth spectrum components may be removed by filtering\citep{Parsons:2012ke} or subtraction \citep{Bowman:2009vc,Liu:2009uy,Dillon:2014}, both employing the separation of the foregrounds and signal in the Fourier domain.

We will focus on the fluctuating signal, assuming that the smooth spectrum components of the foregrounds and background sources are properly avoided and/or subtracted. The effective differential brightness temperature now includes a contribution from the forest absorption features. 
\begin{equation}\label{eq:dtb_prime}
\dtb \to \dtb'  \approx \dtb   - T_{f}
\end{equation}
where $T_f = T_{RL} \tau_{21}$ is the ``forest temperature". We can see how the power spectrum is transformed by the inclusion of $T_f$ by inserting Equation (\ref{eq:dtb_prime}) into Equation (\ref{eq:ps}) 
\begin{equation} \label{eq:ps_with_RL}
  P_b \to P_b' =  P_b  +  P_f- 2 \text{Re}(P_{f,b})
 \end{equation}  
Where $P_b \equiv P_{\dtb}$, $P_b' \equiv P_{\dtb'}$, $P_f \equiv P_{T_f}$ and $P_{f,b} \equiv P_{f,\dtb}$. Equation (\ref{eq:ps_with_RL}) sums up how the forest modifies the power spectrum that we expect to observe in upcoming 21 cm observations. Essentially, smooth spectrum power from $T_{RL}$ is leaked from the largest spatial modes to those occupied by  $\dtb$ via a convolution with the power spectrum of the optical depth field. The magnitude of this leakage will increase with the magnitude of the optical depth.
\subsection{The Morphology of the Forest Power Spectrum}
	The first thing one might ask concerning the forest contribution described in Equation (\ref{eq:ps_with_RL}) is how the magnitudes of the two contributions compare to each other and what their qualitative features are. While we will answer these questions with simulations it is useful to gain as much insight as we can through analytic methods. We start with $P_f$ which can be decomposed (see Appendix \ref{app:compare_pf} for a derivation) into a sum of auto power spectra $P_j$ originating from each individual RL source behind or within an imaged volume of IGM and their cross power spectra, $P_{j,k}$. 
	\begin{equation} \label{eq:sum_forest_sources}
 P_f  = \frac{1}{V} \left \langle \left| \widetilde{T_{RL} \tau_{21}}  \right|^2  \right \rangle = \displaystyle \sum_j      P_j +2\text{Re}\left(   \displaystyle \sum_{j < k }  P_{j,k} \right)
\end{equation}	
If all of the background sources are unresolved\footnote{a fair assumption given the large synthesized beams of interferometers and small angular extent of high redshift sources} then each $P_j$ is the absolute magnitude of the Fourier transform of a function that is a delta function in the perpendicular to LoS directions. As a result, each $P_j$ in Equation (\ref{eq:sum_forest_sources}) is constant in $k_\perp$. The cross multiplying $P_{j,k}$ terms are not so simple; however, we show in Appendix \ref{app:compare_pf} that in the absence of clustering, the cross sum only contributes to $P_f$ at the $10\%$ level for $\kpara \gtrsim 0.1 \Mpci$. At these scales, $P_f$ only has considerable structure along $\kpara$ 
\begin{equation}\label{eq:pf_simple}
P_f({\bf k}) \approx \sum_j P_j(\kpara) = \frac{D_M^2 \lambda^4}{4 k_B^2 \Omega_{cube}} P_{\tau_{21}}^{LoS}(k_\parallel)\langle \sum_j s^2_j \rangle
\end{equation}
where $\lambda=\lambda_{21}(1+z)$ is the observed wavelength of 21 cm light emitted from the center of the imaged volume, $D_M$ is the comoving distance to the data cube, and $\Omega_{cube}$ is the solid angle subtended by the cube. In the second step, we have expressed each $P_j$ in terms of the flux of each source, $s_j$, and the 1D power spectrum along the line-of-site to that source, $P_{\tau_{21}}^{LoS}$. In addition $P_f$ is positive so that it will always add to the power spectrum amplitude 

We can convert the sum in Equation (\ref{eq:pf_simple}) to an integral over the radio luminosity function
\begin{equation}\label{eq:integral}
 P_f \approx \frac{c D_M^2 \lambda^4}{4 k_B^2} P_{\tau_{21}}^{LoS}(\kpara)  \int_z^\infty \int_0^\infty s'^2 \rho(z,z',s') \frac{D_M^2(z')}{H(z)} dz'ds' 
\end{equation}
where $\rho(z,z',s') \equiv \frac{dN}{ds' dV_c}$ is the differential number of radio loud sources per comoving volume at redshift $z'$ per flux bin at observed frequency $\nu_{21}/(1+z)$ and $s'$ is the flux at $\nu = \nu_{21}/(1+z)$. 

Equation (\ref{eq:integral}) tells us that the amplitude of the forest power spectrum is set by the integral over the high redshift radio luminosity function multiplied by the average optical depth squared\footnote{By our definition, the power spectrum is the Fourier transform squared of $\Delta \tau_{21}$, not $\delta_{\tau_{21}} = \Delta \tau_{21}/\langle \tau_{21} \rangle$ which is often used in other work. Hence our power spectrum amplitude is set by $\langle \tau_{21} \rangle^2$} while the shape of the forest power spectrum is set by the 1D LoS power spectrum of optical depth fluctuations.

$P_{f,b}$ does not separate so conveniently but we can gain insight into whether it adds or subtracts to Equation (\ref{eq:ps_with_RL}) by considering the  physical phenomena that govern $T_f$ and $\dtb$. Expanding Equation (\ref{eq:dtb_no_RL}) and $T_f$ in terms of the IGM properties using Equation (\ref{eq:tau}) one can see that $P_{f,b}$ is the cross power spectrum between the two quantities:
	\begin{equation}\label{eq:dtb_full}
	\dtb \approx 9 x_{HI} (1+\delta)(1+z)^{1/2} \left[ 1 - \frac{T_{CMB}}{T_s} \right] \left[ \frac{H(z)}{dv_\parallel/dr_\parallel} \right] \mK
	\end{equation}
	and
	\begin{equation}\label{eq:tf_full}
	T_f  \approx 0.009 x_{HI} (1+\delta)(1+z)^{1/2}\frac{T_{RL}}{T_s}\left[ \frac{H(z)}{dv_\parallel/dr_\parallel} \right]
	\end{equation} 
	
	Before the reionization era, $x_{HI}$ is relatively homogenous so that fluctuations in $\dtb$ are governed primarily by those in $T_s$. Regions of the IGM with larger $T_s$ will have more positive $\dtb$ but smaller $T_f$. Because of this anti-correlation between $T_f$ and $\dtb$, $\text{Re}(P_{f,b})$ is negative during the pre-reionization era and the net effect will be for it to increase the power spectrum amplitude through its negative contribution in Equation (\ref{eq:ps_with_RL}). At lower redshifts, after X-rays have heated the IGM, $T_s \gg T_{CMB}$, and $\dtb$ becomes independent of $T_s$. As a result, $\dtb$ is always positively correlated with $x_{HI}$ as is $T_f$. $\text{Re}(P_{f,b})$ is positive with a net effect of subtracting from the power spectrum amplitude. We are unable to make any more progress analytically, but we will reexamine the cross power term in our simulation results below.
	
	We now move on to describe our simulations. We will return to our discussion of the power spectrum morphology in the context of our simulation results in Section \ref{sec:results}.


 \vskip 1.0 cm
\section{ SIMULATIONS}\label{sec:sim}

In this section we describe the semi-numerical simulations that we use to explore a range of IGM thermal histories along with the the semi-empirical RL source model that we employ to add the 21 cm forest signal. 

\subsection{Simulations of the Optical Depth of the IGM}

Our IGM simulations are run using a parallelized version of the public, semi-numerical 21cmFAST code\footnote{\url{http:/homepage.sns.it/mesinger/Sim}} described in \citet{Mesinger:2011tb}. Tests of the code can be found in \citet{Mesinger:2007hl,Zahn:2011,Mesinger:2011tb}. The simulation box is 750 Mpc on a side, with resolution of $500^3$. Different scenarios for $\tau_{21}$ can be obtained by exploring histories of the spin temperature, $T_s$ and/or the neutral fraction, $x_{HI}$.

 21cmFAST includes sources of both UV ionizing photons and X-rays.  The former dominate reionization (i.e. $x_{HI}$), except for extreme scenarios we do not consider in this work \citep{Furlanetto:2006, McQuinn:2012, Mesinger:2013vm}.
  Since a full parameter study is beyond the scope of this work, and since the bulk of the relevant signal is likely during the pre-reionization epoch,  we fix the ionizing emissivity of galaxies (and hence the reionization history), to agree with the Thompson scattering optical depth from WMAP \citep{Komatsu:2011in}.  Instead we focus on the X-ray emissivity and its impact on $T_s$.

 $T_s$ is affected by a variety of processes. These include Ly-$\alpha$ photons which couple to the hyperfine transition through the Wouthuysen-Field effect \citep{Wouthuysen:1952hd,Field:1958}, particle collisions, and emission or absorption of CMB photons. The coupling of $T_s$ to these processes is described by (e.g. \citealt{Furlanetto:2006tc}):
 \begin{equation}
 T_s^{-1} = \frac{T_{CMB}^{-1}+x_c T_k^{-1} + x_\alpha T_c^{-1}}{1+x_c + x_\alpha}
 \end{equation}
 where $T_k$ is the kinetic temperature of the HI gas, $T_c$ is the color temperature of Ly-$\alpha$ photons, and $x_c$ and $x_\alpha$ are the collisional and Ly-$\alpha$ coupling constants. Due to the high optical depth of the neutral IGM to Ly-$\alpha$ photons, the color temperature is very closely coupled to the kinetic temperature, $T_c \approx T_k$  \citep{Wouthuysen:1952hd,Hirata:2006} .

Although the self-annihilation of some dark matter candidates can contribute significantly \citep{Valdes:2013}, in fiducial models $T_k$ is predominantly determined by X-ray heating (e.g. \citealt{Furlanetto:2006tc}). Hence, we explore a range of optical depth histories by running simulations for different galactic X-ray emissivities.

We use the fiducial model of X-ray heating described in \citet{Mesinger:2013vm}, adopting a spectral energy index of $\alpha=1.5$ and an obscuration threshold of 300 eV. We parameterize the X-ray luminosity by a dimensionless efficiency parameter, $f_X$.  Our fiducial model, $f_X=1$ corresponds to 0.2 photons per stellar baryon, or a total X-ray luminosity above $h\nu_0=0.3$ keV of $L_{\rm X, {\rm 0.3+keV}}\approx 10^{40}$ erg s$^{-1}$ ($\Msun$ yr$^{-1})^{-1}$. This choice is consistent with (a factor of $\sim$2 higher than) an extrapolation from the 0.5--8 keV measurement of \citet{Mineo:2013} that $L_{\rm X, {\rm 0.5-8keV}}\approx3\times10^{39}$ erg s$^{-1}$ ($\Msun$ yr$^{-1})^{-1}$. 

\begin{figure}
\includegraphics[width=.5\textwidth]{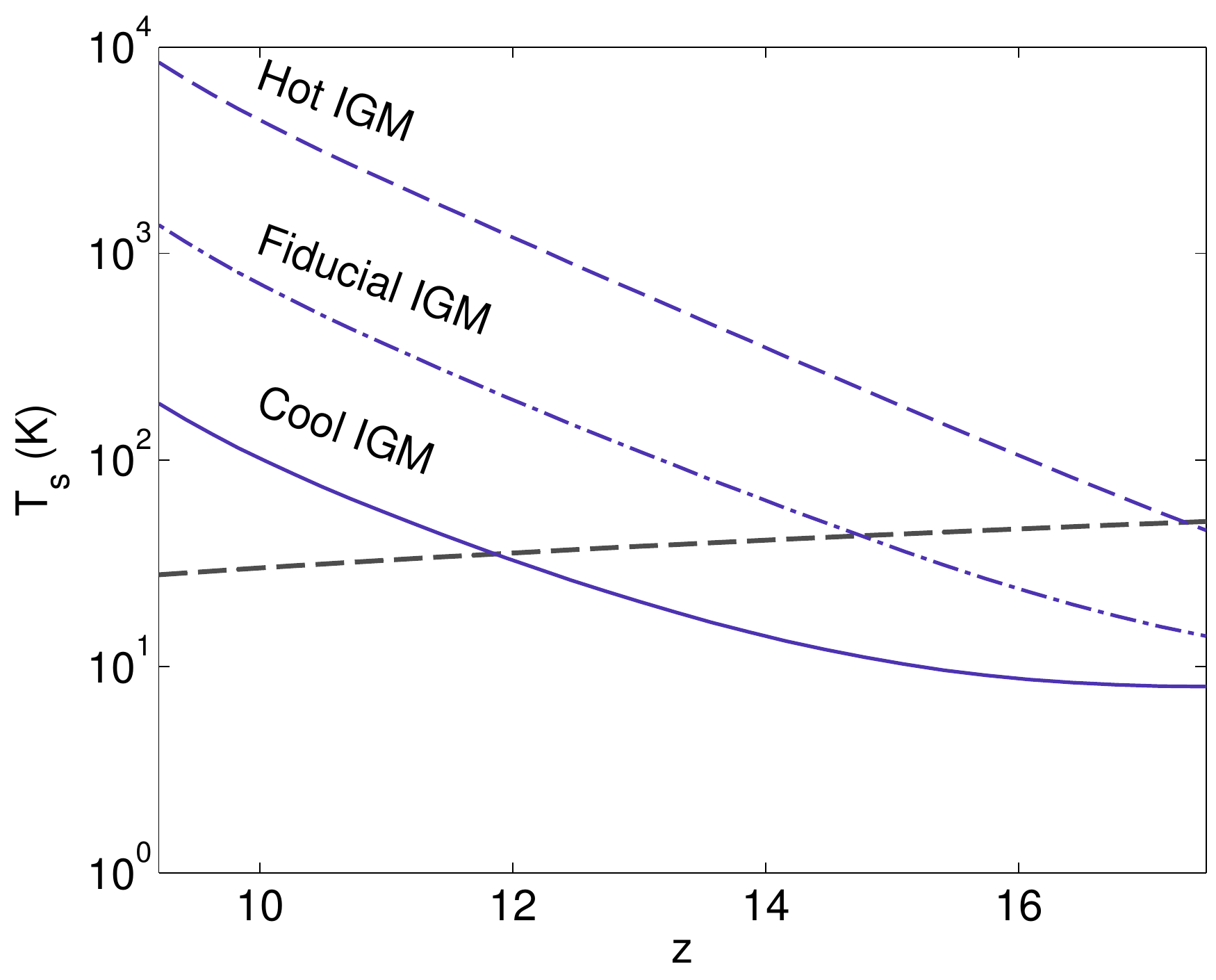}
\caption{The mean thermal evolution of our IGM simulations for our three models. ``cool IGM''- solid lines, ``fiducial IGM''- dashed-dotted lines, and ``hot IGM'' - dashed lines. $\langle T_s \rangle$ is plotted in lavendar. Varying $f_X$ effectively shifts $\langle T_s \rangle$ in redshift.} 
\label{fig:Thermal_Evolution}
\end{figure}

Summarized in Table \ref{tab:models} are our three values of $f_X$: a ``fiducial IGM" model with $f_X=1$ corresponding to the fiducial value in \citet{Mesinger:2013vm}, a ``hot IGM" model with $f_X = 5$, and a ``cool IGM" model with $f_X = 0.2$. In Figure \ref{fig:Thermal_Evolution} We show the evolution of the mean spin and brightness temperatures from our simulations. Over the range of emissivities considered, the effect of varying $f_X$ is to shift the evolution of $\langle T_s \rangle$ in redshift.  Because $P_f$ varies as $\langle \tau_{21} \rangle^2 \sim \langle T_s \rangle^{-2}$ and $f_X$ simply shifts $\langle T_s \rangle$ in redshift, this relatively modest spread in $f_X$ is sufficient to understand a broader range of expected outcomes, as we shall see below.

\begin{table}
\begin{center}
\caption{IGM Heating Parameters}
\begin{tabular}{cc}   \hline\hline
Number Name & $f_X$ \\ \hline
 Hot  IGM & $5.0$ \\ 
 Fiducial IGM & $1.0$ \\
 Cool IGM & $0.2$ \\ \hline
\end{tabular}
\label{tab:models}
\end{center}
\end{table}

\subsection{The Model of the Radio Loud Source Distribution} 
We now review present constraints on the RL source distribution and describe the semi-empirical radio luminosity function that we use to simulate the global 21 cm forest. To gain perspective of how our choice of population model might compare to other theoretical work we determine which flux ranges are relevant to the sum in Equation (\ref{eq:pf_simple}) and compare the counts of sources in W08 to those in H04.  We also describe our method for combining the simulated radio sources with our simulations of the IGM. 

\subsubsection{Review of Constraints and Predictions of High Redshift Radio Counts}

Constraints on the luminosity function of the most luminous radio loud sources are presently limited to $z \sim 4$ \citep{DeZotti:2010} Confirmed in these works, is that the comoving density of ultra steep spectrum sources peaks at $z \sim 2$ with little evidence for evolution out to $z \gtrsim 4.5$.

 To model the abundance of RL quasars with $6 \lesssim z \lesssim 20$ one must rely on theoretical extrapolations.  \citet{2004ApJ...612..698H} give estimates of source counts by assigning black hole masses to a halo mass function using the black hole mass-velocity dispersion relation of \citet{Wyithe:2003ev}. The RL fraction is derived assuming Eddington accretion, and the RL-i band luminosity correlation observed by \citet{Ivezic:2002}. 

 More sophisticated attempts at predicting the bolometric luminosities of high redshift quasars up to $z=11$ have been undertaken using hydrodynamic simulations with self consistent models for black hole growth and feedback \citep{DeGraf:2012ta}. Even with a more nuanced treatment of the luminosity distribution, the RL fraction at high redshift still remains a wide open question.  Indeed, the purpose of this work is to propose a technique for determining this population by showing that an empirically motivated RL population can have significant and observable features in the power spectrum for a range of thermal scenarios. 

\subsubsection{Our Choice of Population Model}

 We choose to work with the RL AGN population described in W08 in which sources are generated by sampling extrapolated radio luminosity functions biased to structure from a CDM simulation. Specifically, the radio luminosity function used is that ``Model C." from \citet{Willott:2001} which describes the high and low luminosity populations of AGN as Schechter functions. The redshift evolution of the low luminosity population is modeled as a power law in redshift while the high luminosity component as a gaussian with a mean of $z\approx 1.9$. Lists of source positions, fluxes, and morphologies from the Wilman simulation are downloadable through a web interface\footnote{\url{http://s-cubed.physics.ox.ac.uk/s3_sex}}. 

\begin{figure}
\includegraphics[width=.48 \textwidth]{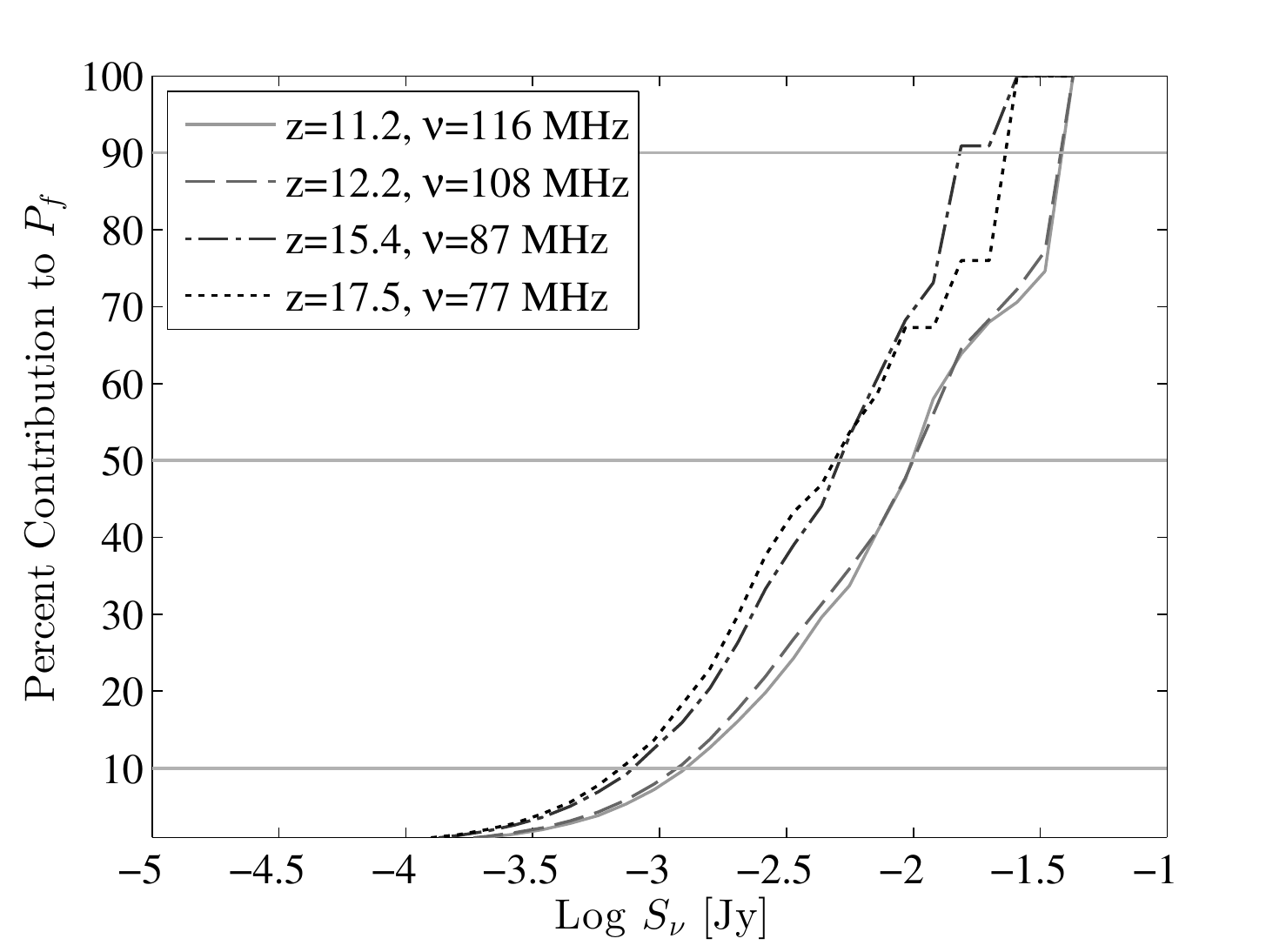}
\caption{The 21 cm forest is dominated by sources in the 1-10 mJy flux range. We plot the sum of fluxes squared, in Equation (\ref{eq:integral}), for $S < S_\nu$.  A detection of $P_f$ would constrain the high redshift source counts at these flux intervals.}
\label{fig:flux_integral}
\end{figure}

Having chosen our population model, we can employ our formalism from Section \ref{sec:theory} to understand which sub population of the luminosity function contributes most to $P_f$.  In Figure \ref{fig:flux_integral}, we plot the percent contribution of sources below a threshold, $S_\nu$, to $P_f$ from the flux squared sum in Equation (\ref{eq:pf_simple}). One can see that roughly 75\% of the contribution to $P_f$ comes from sources with fluxes between $1-10\mJy$ at $80-115 \MHz$. At lower redshifts, the integral curves are increasingly dominated by higher fluxes as the sources with the greatest fluxes increase in number. The detection or lack of detection of the features we find using this simulation would either confirm or reject the W08 model for sources with $S_\nu$ between $1$ and $10 \mJy$.
While this paper is a study of observability for one model, in future work we will determine what range of  RL population this technique can constrain.

It is worth getting an order of magnitude idea of how our choice of the W08 semi-empirical model might compare to other theoretical predictions of the radio luminosity function. In Appendix \ref{app:compare} we compare the source counts in our semi-empirical prediction to the more physically motivated bottom up model in H04. The counts of W08 sources contributing to the bulk of $P_f$  tend to be more numerous than those in H04 by a factor of $\approx 10$ at $z \sim 12$ to $\approx 80$ for $z \sim 15-20$, underscoring the need for a full parameter space study. Even though such a study is beyond the scope of this paper, our extrapolated results in Figure \ref{fig:sn_params} show that the range of populations that the power spectrum can constrain depends heavily on the IGM's thermal history.

\subsection{Adding Sources to the Simulation}

  We simulate the theoretical power spectra accessible to upcoming observations by drawing 36 random sub-fields from the W08 simulations and combining them with 36 random 8MHz slices from our IGM simulations. The number of subfields is chosen to roughly correspond to the $\sim (30\degree)^2$ FoV of the MWA.

While our analytic approach in Section \ref{sec:theory} does not account for sources within the imaged volume, we incorporate them into our simulation by determining the location of DM halos down to masses of $5 \times 10^9 M_\odot$ through the excursion-set + perturbation theory approach outlined in \citet{Mesinger:2007hl}. We then populate these dark matter halos with RL sources, monotonically assigning the most luminous sources at 151 MHz\footnote{We order sources by their luminosity  at {\it observed} frequency of 151 MHz at regardless of their redshift which varies very little over the span of an $8 \MHz$ data cube so that we are approximately comparing their rest frame luminosities.} to the most massive halos. 
 Sources falling behind the cubes retain their original positions. All W08 sources are unresolved in our IGM simulation; hence, for each pixel the fluxes for all sources behind that pixel are summed together to give $S_{pix}$. This flux cube is converted to temperature using the Rayleigh-Jeans equation,
\begin{equation}  
T_{pix} = \frac{ \lambda^2 S_{pix} }{2 k_B \Omega_{pix}},
\end{equation}
where $\Omega_{pix}$ is the solid angle subtended be each simulation pixel\footnote{We show in Appendix \ref{app:compare_pf}, that the choice of pixel solid angle does not effect $P_f$}. Finally we introduce quasar absorption by multiplying this source cube by our $\tau_{21}$ cube $T_f \approx T_{pix} \tau_{21}$.

\section{ SIMULATION RESULTS} \label{sec:results}
In this section we present the results of our combined IGM-RL population model by computing the spherical and cylindrical power spectrum, $P(k)$, averaged over our 36 sub-cubes. We identify the regions of $k$-space in which the forest is dominant and might be used to constrain the high redshift radio luminosity function and discuss the morphology of the observed power spectra, verifying the essential results of Section \ref{sec:theory}.

\subsection{Computing Power Spectra}
Power spectra are computed using a direct Fast Fourier transform of each data cube multiplied by a kaiser window along the LoS with attenuation parameter $\beta = 3.5$. In averaging over bins of our spherical power spectra, we exclude the ``wedge", the region of $k$-space heavily contaminated by foregrounds given by \citep{Vedantham:2013wh,Morales:2012uk}
\begin{equation} \label{eq:wedge}
\kpara \le  \sin \frac{\Theta}{2} \left( \frac{D_M(z)}{D_H} \frac{E(z)}{(1+z)} \right) \kperp
\end{equation}
where $z$ is the redshift of a data cube's center frequency,  $D_M(z)$ is the comoving distance, $E(z)=H(z)/H_0$, and $\Theta$ is the FWHM of the primary beam which we calculate using a short dipole model of the MWA antenna element.  Table \ref{tab:mwa_params} gives the FWHM value of our primary beam model for several different frequencies. 

\begin{figure*}
\includegraphics[width=\textwidth]{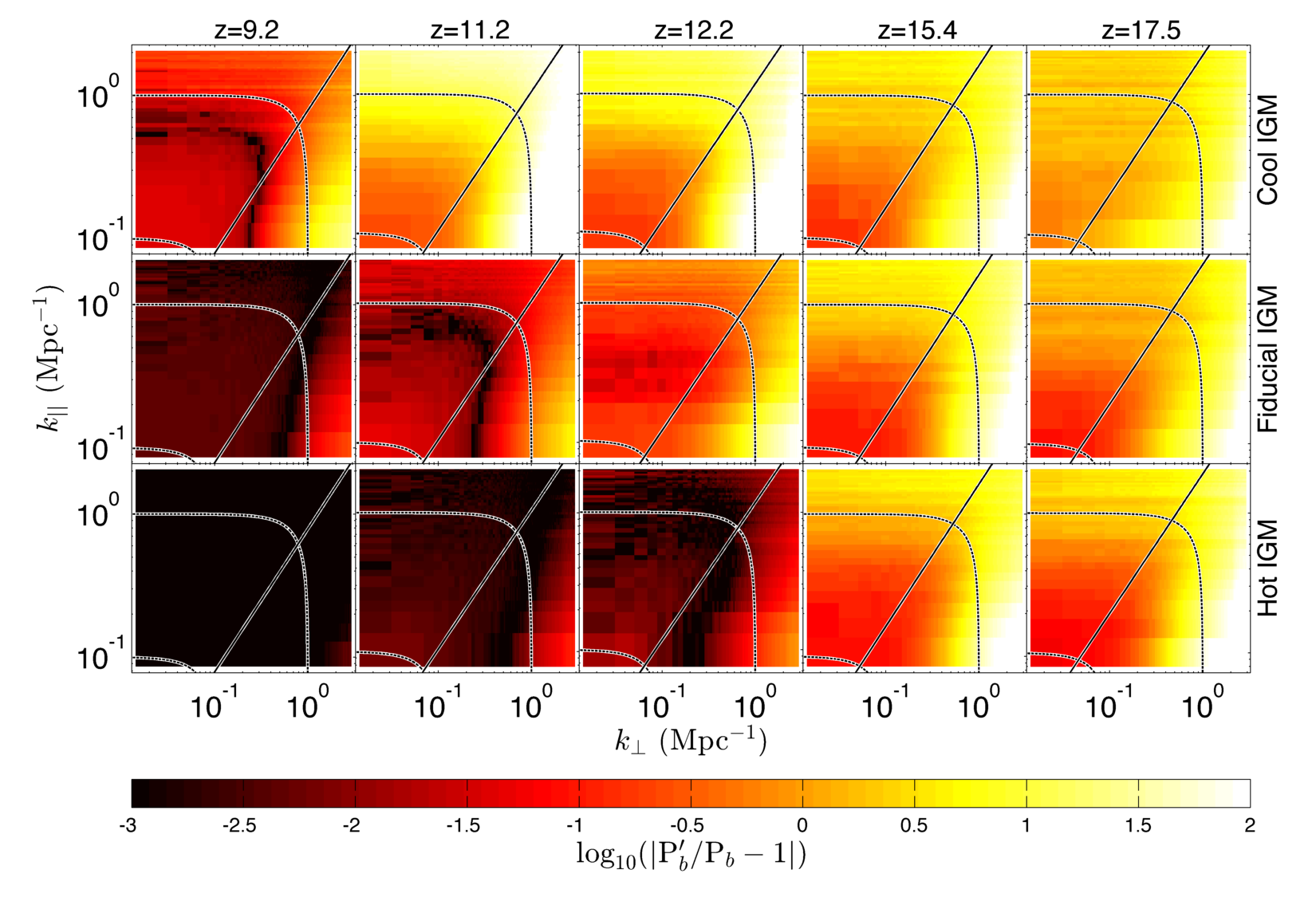}
\caption{For every heating scenario we study, there is some redshift and region within the EoR window for which 21 cm forest dominates the power spectrum. Here we show the fractional difference between the power spectrum with, ($P_b'$), and without ($P_b$) the forest for the redshifts (top to bottom) 9.2, 11.2, 12.2, 15.4, and 17.5. The diagonal lines denote the location of the ``wedge".  By $z \gtrsim 12.2$ there is a substantial region  ($\kpara \gtrsim 0.5 \Mpci$) of  the Fourier volume that our simulations cover in which the forest dominates $P_b$ by a factor of a few.}\label{fig:forest_compare} 
\end{figure*}

\subsection{Simulation Output and the Location of the Forest in $k$-space}

We now discuss the power spectra output by our simulations and  the significant features produced by the forest. 

To isolate the the effect of the forest and to compare its significance to the brightness temperature power spectrum, $P_b$, we plot the fractional difference between $P_b'$, the power spectrum with the forest ,and $P_b$ in Figure \ref{fig:forest_compare}. We see that the forest introduces a significant feature, especially at the smallest scales. This feature is most prominent at high redshifts and less emissive heating models, when the IGM is cool. For our cool model, the forest feature dominates $P_b$ by over a factor of 100 for a wide range of redshifts. In the fiducial model, the dominant region is primarily at larger values of $\kpara$, though dominance by a factor of a few is visible at $z=12.2$ and $z=17.5$. In our hot model, a significant feature is visible only for $z \gtrsim 12.2$.

 For all heating scenarios, there are redshifts $z \gtrsim 12.2$ in which the same region of Fourier space contains a strong forest signal that dominates $P_b$ by a factor of at least a few.  Fortunately for those interested in the brightness temperature signal, the region $k \lesssim 10^{-1} \Mpci$ remains dominated by $P_b$. Hence at pre-reionization redshifts,  $k \lesssim 10^{-1} \Mpci$ can still be used to constrain cosmology and the thermal history of the IGM. With the thermal properties of the IGM determined, one may constrain  the high redshift RL population using the forest power spectrum signal at $k \gtrsim 0.5 \Mpci$.

\begin{figure*}
\centering
\includegraphics[width=\textwidth]{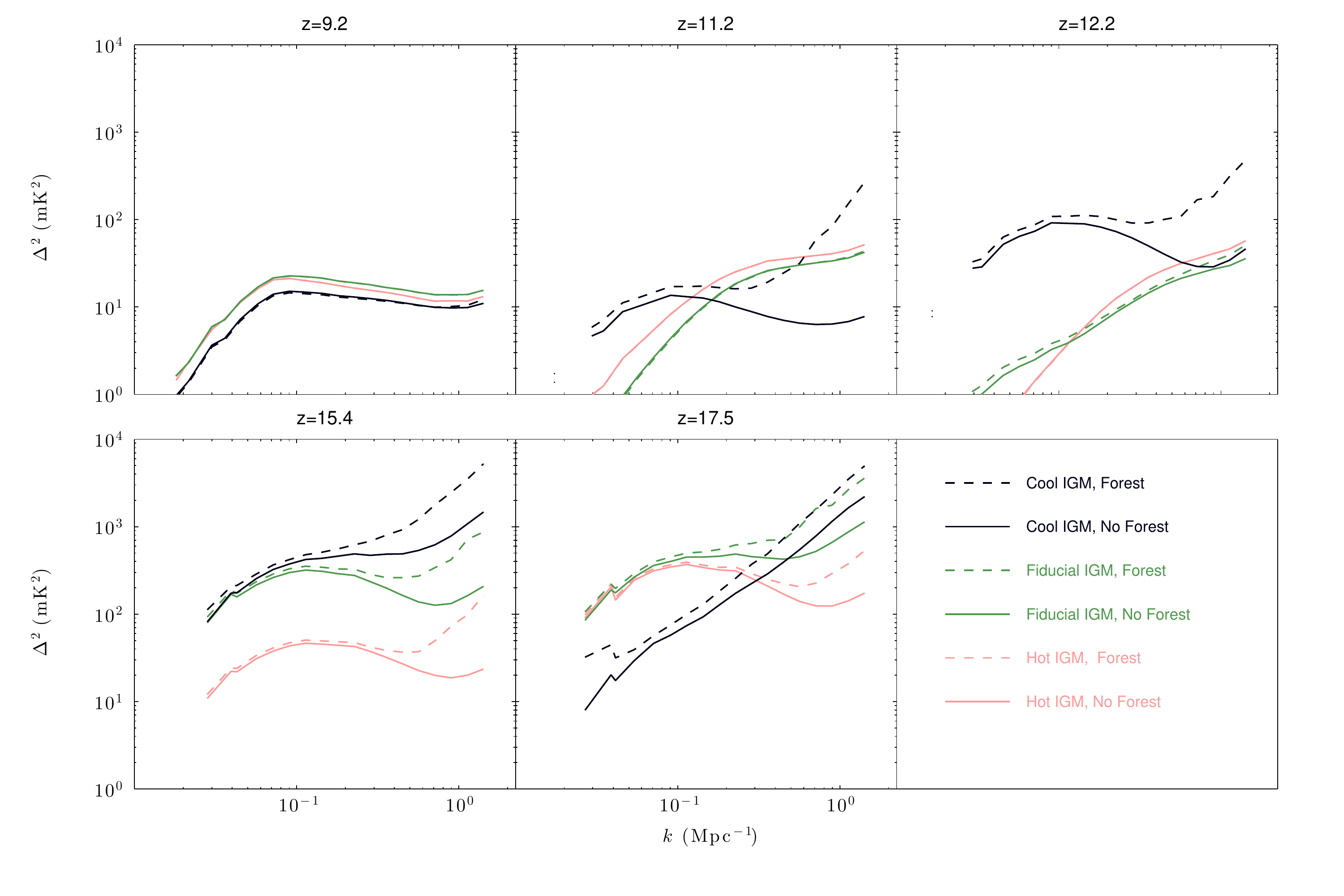}
\caption{The 21 cm forest dominates the spherically averaged power spectrum for $k \gtrsim 0.5 \Mpci$. Plotted is the spherically averaged power spectrum with  (dashed lines) and without (solid lines) the presence of the 21 cm forest. In our cool model, the forest causes a significant power increase at $k \gtrsim 0.5 \Mpci$ at redshifts as low as $z = 11.2$.   At $z=15.4$ we see a significant feature in all thermal scenarios. Our cool IGM model experiences a reduction in the power spectrum amplitude at $z \gtrsim 17.5$ as it passes through the X-ray heating peak.  }
\label{fig:ps}
\end{figure*}

The first generation of interferometers will not be sensitive enough to measure the cylindrical power spectrum with high S/N but will rather measure the spherically averaged power spectrum. We compute spherically averaged power spectra from data cubes with and without the presence of forest absorption and excluding the wedge. We plot these power spectra in Figure \ref{fig:ps}. In all of the heating scenarios considered, the forest introduces significant power at $k \gtrsim 0.5\Mpci$ for $z \gtrsim 15.4$. Hence, it is in principle possible to constrain the distribution of RL AGN at high redshift for a range of heating scenarios.

We note that the high-$k$ region extends into our simulations' Nyquist frequency of $2.1 \Mpci$. We ensure that the  forest dominance is not an aliasing effect by running simulations on a 125 Mpc cube with six times higher resolution.  The results in the the overlapping $k$-space regions agree well with these larger volume, lower resolution simulations.

\begin{figure*}
\centering
\includegraphics[width=\textwidth]{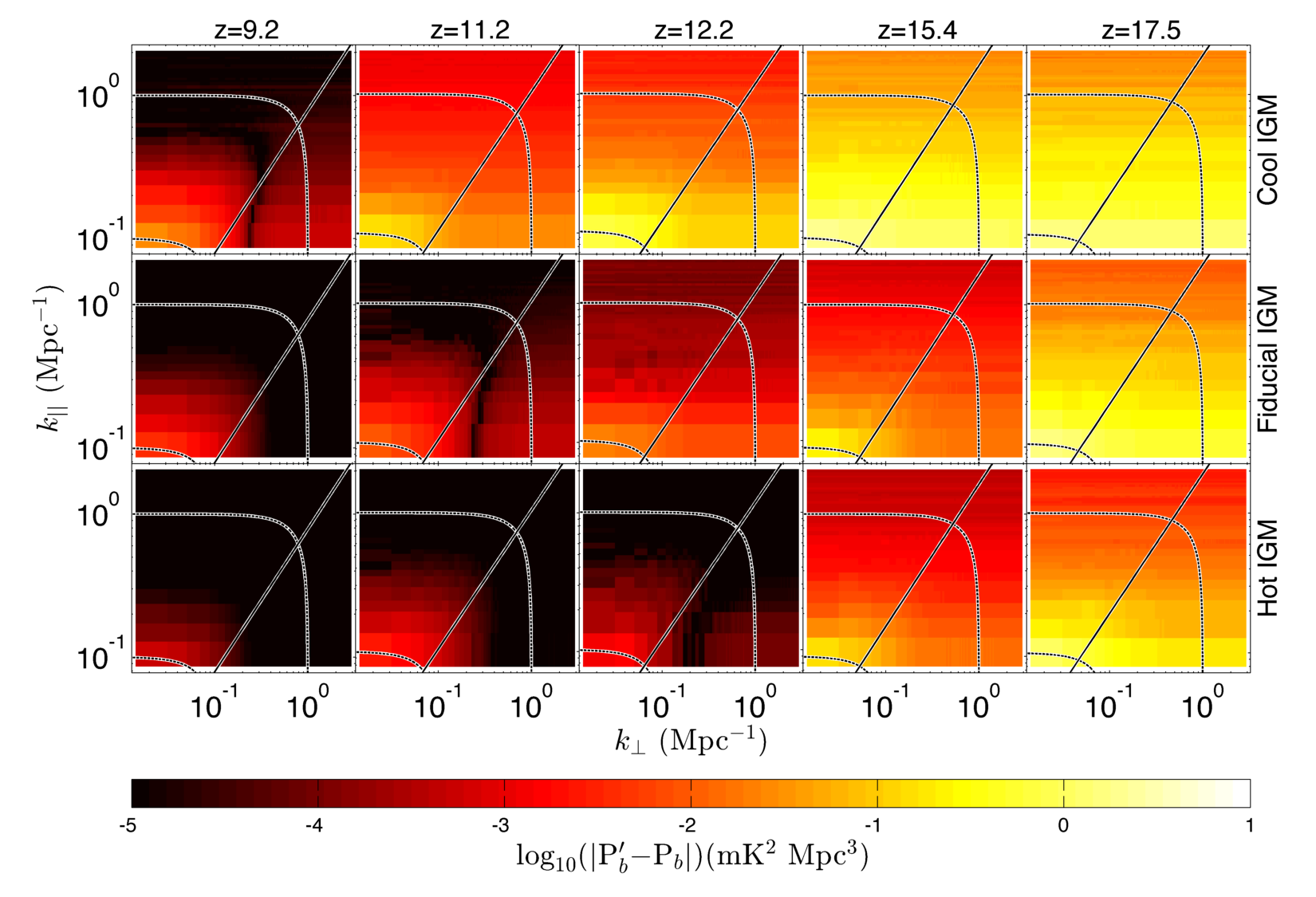}
\caption{We plot the magnitude of the difference between the 21 cm  power spectrum with and without the presence of the 21 cm forest including the auto-power and cross power terms of Equation (\ref{eq:ps_with_RL}). At high redshifts and low $f_X$, there is little $\kperp$ structure in $P_b'-P_b$, indicating that $P_f$ is the significant contributer. At lower redshifts and higher $f_X$, we see signficant $\kperp$ structure, indicating that in a heated IGM, $P_b' - P_b$ is dominated by $P_{f,b}$ which is somewhat spherically symmetric and negative at large $k$. The trough in the low redshift plots marks the region where $P_f-2\text{Re}(P_{f,b})$ transitions from negative (for small k) to positive (for large k). }
\label{fig:ps_modified}
\end{figure*}

\subsection{The Morphology of the Simulation results.}
We now explain the morphology of our simulation results and verify our analytic predictions in Section \ref{sec:theory}. 

We noted in Figure \ref{fig:forest_compare} that the 21 cm forest dominates the power spectrum both at large $\kperp$ and $\kpara$. The former observation is consistent with a forest power spectrum that is uniform in $\kperp$. In Figure \ref{fig:ps_modified} we show $|P_b'-P_b|$ and see that at high redshift and cool heating models, the forest power spectrum is mostly uniform in $\perp$ though at lower redshifts and hotter IGM, there is significant $\kperp$ structure. Since in section 
\ref{sec:theory} we showed that $P_f$ only varies along $\kpara$, this suggests that the cross power spectrum, $P_{f,b}$ is the prime contributor to $P_b' - P_b$ in a hot IGM, while $P_f$ is in a cool one. The trough at lower redshifts, at $k \sim 0.5 \Mpci$ is caused by the fact that $-2P_{f,b}$ is negative as we shall see below.

A potentially interesting consequence of the auto-terms invariance in $\kperp$ is a potential for contaminating the separation of powers analysis advocated in \citet{Barkana:2005bi}. We may Taylor expand $P_f$  

\begin{equation}
P_f(\kpara) = P_f(k \mu) = \sum_{n=1}^{\infty} \frac{1}{n!}\frac{\partial P_f}{\partial (\mu k)}_{\mu k = 0} ( \mu k)^n,
\end{equation}

so $P_f$ introduces signal over a wide range of powers of $\mu$ and has the potential to contaminate the cosmological $\mu^4$ and $\mu^6$ components of the brightness temperature power spectrum. On the other hand, the small $k$, where the perturbative expansion is most accurate, is dominated by the diffuse brightness temperature emission. In all but the coolest heating models, contamination will likely be small, since we can see in Figure \ref{fig:forest_compare} that $P_f \lesssim 0.1 P_b$ at $k \lesssim 0.1 \Mpci$. 

Decomposing the forest signal into powers of $\mu$ may be another way of distinguishing it from the brightness temperature. Even within the ``IGM dominated" region. Detailed analysis on contamination of the cosmological signal and additional distinguishability offered by the angular dependence is beyond the scope of this paper will be the subject of future work. 

To be more quantitative, we turn our attention to right hand side of Equation (\ref{eq:pf_simple}) and verify our decomposition of the forest power spectrum into $P_{\tau_{21}}^{LoS}$ and the sum of background source fluxes. To do this, we find the summed squares of the fluxes (at the center frequency of the observation) of all sources falling in or behind our data cubes at several redshifts, multiply by the 1D LoS power spectrum of $\tau_{21}$ and compare with $\Delta^2_f$ computed from our simulation as outlined above. We find that Equation (\ref{eq:pf_simple}) consistently underpredicts the simulation amplitude by a factor of 2. However, when we remove the clustering of sources by randomly assigning source positions (rather than using the dark matter biased positions), Equation (\ref{eq:pf_simple}) agrees with simulation output within $5-20\%$ over the studied redshifts. Hence we rewrite Equation (\ref{eq:integral}) as 
\begin{equation}\label{eq:integral_fix}
 P_f \approx A_{cl} \frac{c D_M^2 \lambda^4}{4 k_B^2} P_{\tau_{21}}^{LoS}  \int_z^\infty \int_0^\infty s^2 \rho(z,z',s)  \frac{D^2_M(z')}{H(z')}  dz'ds 
\end{equation}
Where $A_{cl}$ is a constant of order unity that accounts for the boost in power due to clustering. We briefly explain this power boost in Appendix \ref{app:compare_pf}. In Figure $\ref{fig:mod_compare}$ we show the power spectrum, $\Delta^2_f$ computed from our simulation and the prediction from Equation (\ref{eq:pf_simple}) for several redshifts in our fiducial heating model. For $k \gtrsim 10^{-1} \Mpci$, Equation (\ref{eq:pf_simple}) agrees with our simulation at the 10\% level, indicating that we can ignore the cross terms in Equation (\ref{eq:sum_forest_sources}) and consider the forest power spectrum as the simple product of the 1D $\tau_{21}$ power spectrum and the integrated radio luminosity function. 

\begin{figure}
\includegraphics[width=.5 \textwidth]{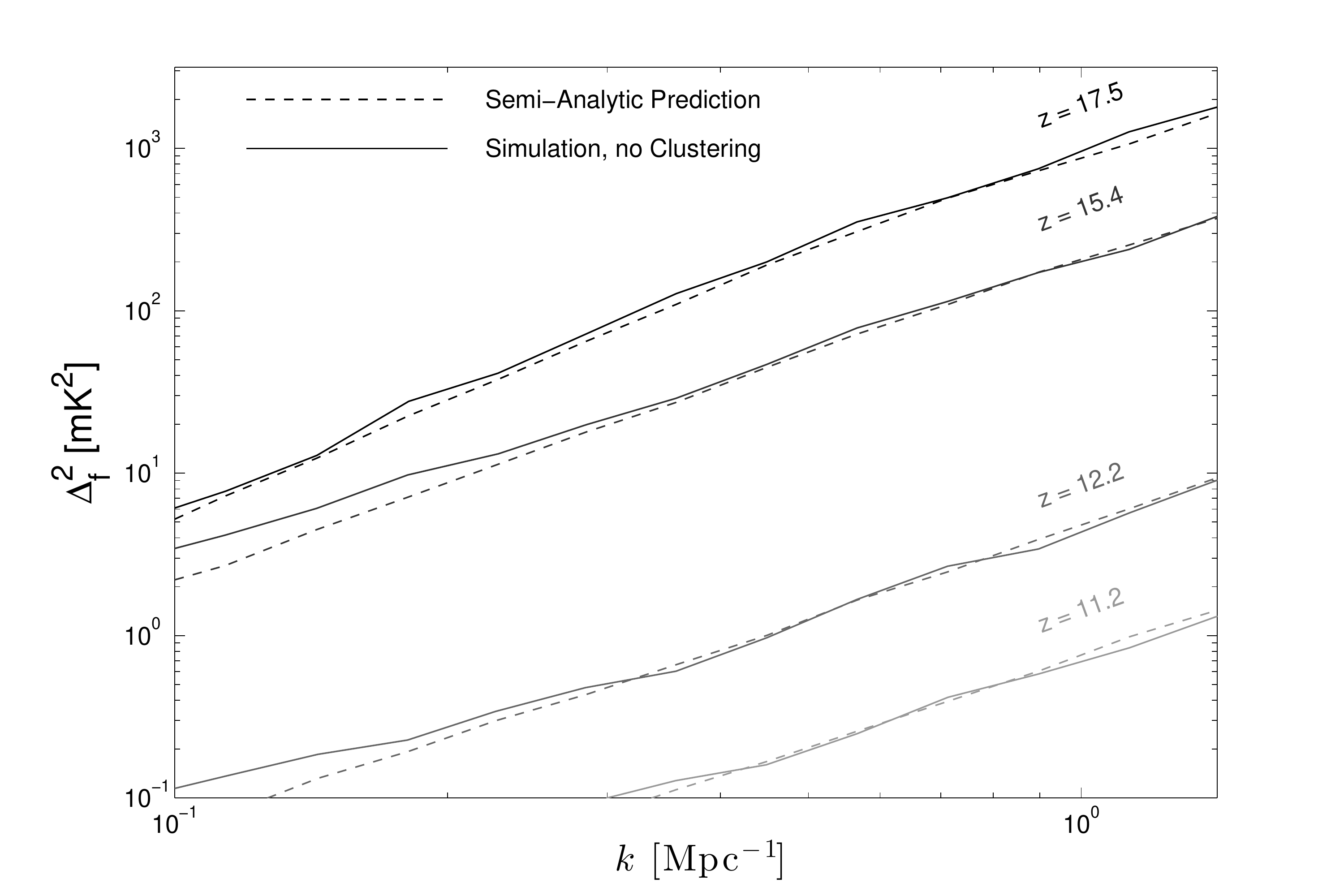}
\caption{Our semi-analytic prediction agrees well with unclustered simulation results. The semi-analytic prediction of Equation (\ref{eq:pf_simple}) is plotted with dashed lines and $\Delta^2_f(k)$ computed directly from our simulation without clustering in solid lines. This demonstrates that for $k\gtrsim 10^{-1} \Mpci$, the cross terms in Equation (\ref{eq:sum_forest_sources}) may be ignored and $P_f$ may be well approximated by the LoS power spectrum of $\tau_{21}$ multiplied by the summed squared fluxes for sources lying in and behind the data cube.}
\label{fig:mod_compare}
\end{figure}

\begin{figure}
\includegraphics[width=.5\textwidth]{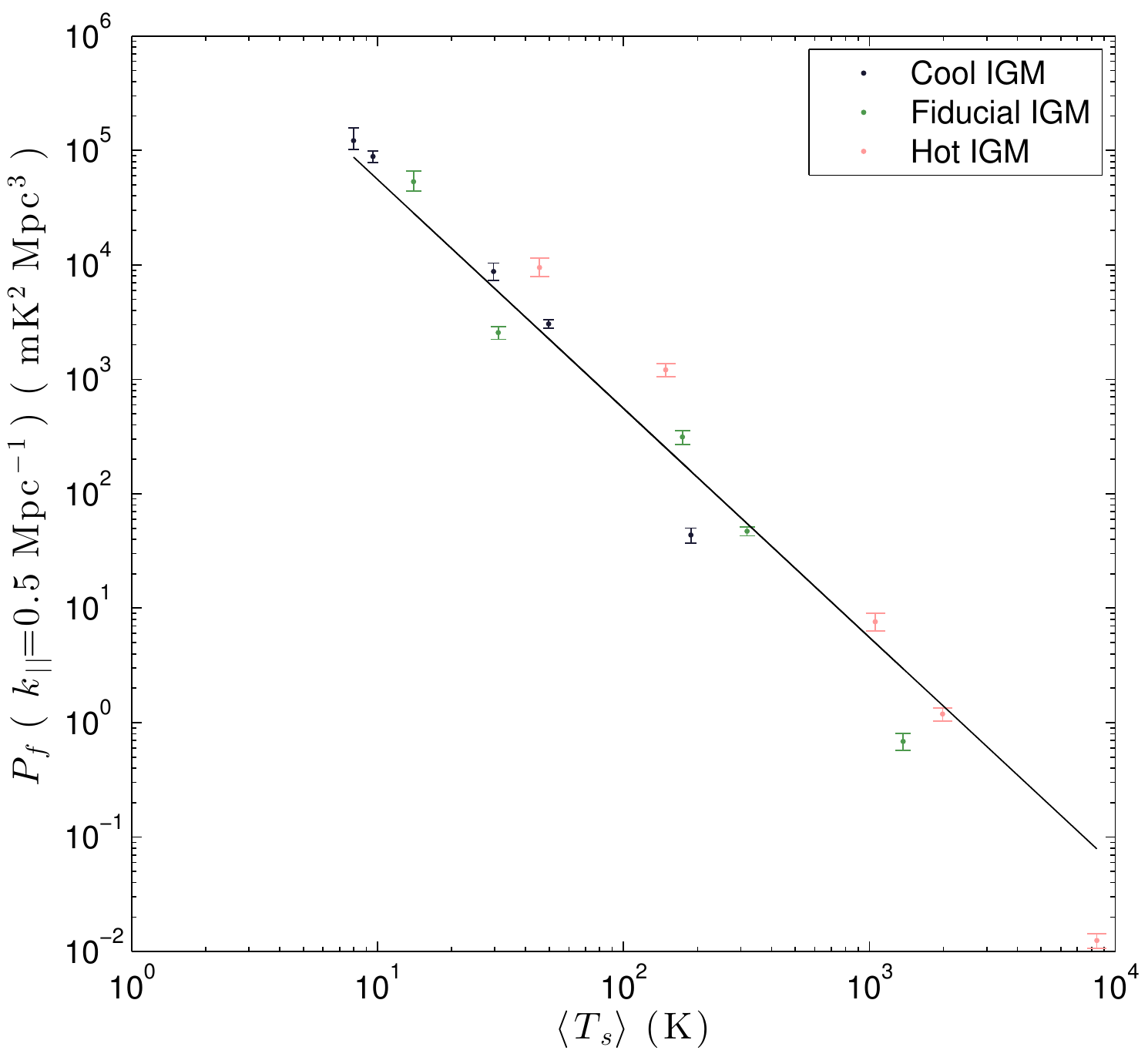}
\caption{ We see that for a fixed quasar distribution, the magnitude of $P_f$ can be parameterized by $\left \langle T_s \right \rangle$ and that the amplitude is consistant with a simple power law. Here, we plot $P_f(k_\parallel)$ at $k_\parallel=0.5 \Mpci$ vs. $\langle T_s \rangle$ for all considered redshifts and $f_X$. The black line is the power law $\langle T_s \rangle^{-2}$ as one might expect for an amplitude set by  $\langle \tau_{21} \rangle^2$ (Equation (\ref{eq:integral})). Inasmuch of this simple trend, a modest spread in heating models gives us a decent understanding of the behavior of the amplitude for $P_f$. This relation holds for the quasar population considered here because the integral over the luminosity function does not change significantly over the redshifts we consider.}
\label{fig:ps_modifiedVsTs}
\end{figure}

A striking feature of Figure \ref{fig:ps_modified} is the apparent similarity of $P_f$ along diagonal sets of different redshifts and models. For example, the ``Cool IGM" model at $z=12.2$ is very similar to the ``Fiducial IGM" result at $z=15.4$ and the ``Hot IGM" at $z=17.5$. It is suggestive that one can obtain the results of one particular thermal model by simply shifting another model in redshift, this translational invariance in redshift demonstrates that we may not need to simulate a broad range of heating models to understand the evolution of the forest power spectrum. Indeed, given our decomposition in Equation (\ref{eq:pf_simple}) where the amplitude of $P_f$ is proportional to $\langle \tau_{21} \rangle^2 \propto \langle T_s^{-1} \rangle^2$, we should expect $\langle T_s \rangle$ to be a more generally applicable parameterization than $f_X$ and redshift during the pre-reionization epoch.
 To show the importance of $\langle T_s \rangle$ as a parameter, we plot, in Figure \ref{fig:ps_modifiedVsTs}, the amplitude of $P_f$ at $k_\parallel = 0.5~\Mpci$ as a function of $\langle T_s \rangle$ for our three heating scenarios and redshifts. Across all thermal models and redshifts, the amplitude of $P_f$ is well described by a power law of $\langle T_s \rangle^{-2}$, consistent with the normalization predicted in Equation (\ref{eq:pf_simple}).

\begin{figure*}
\centering
\includegraphics[width=  \textwidth]{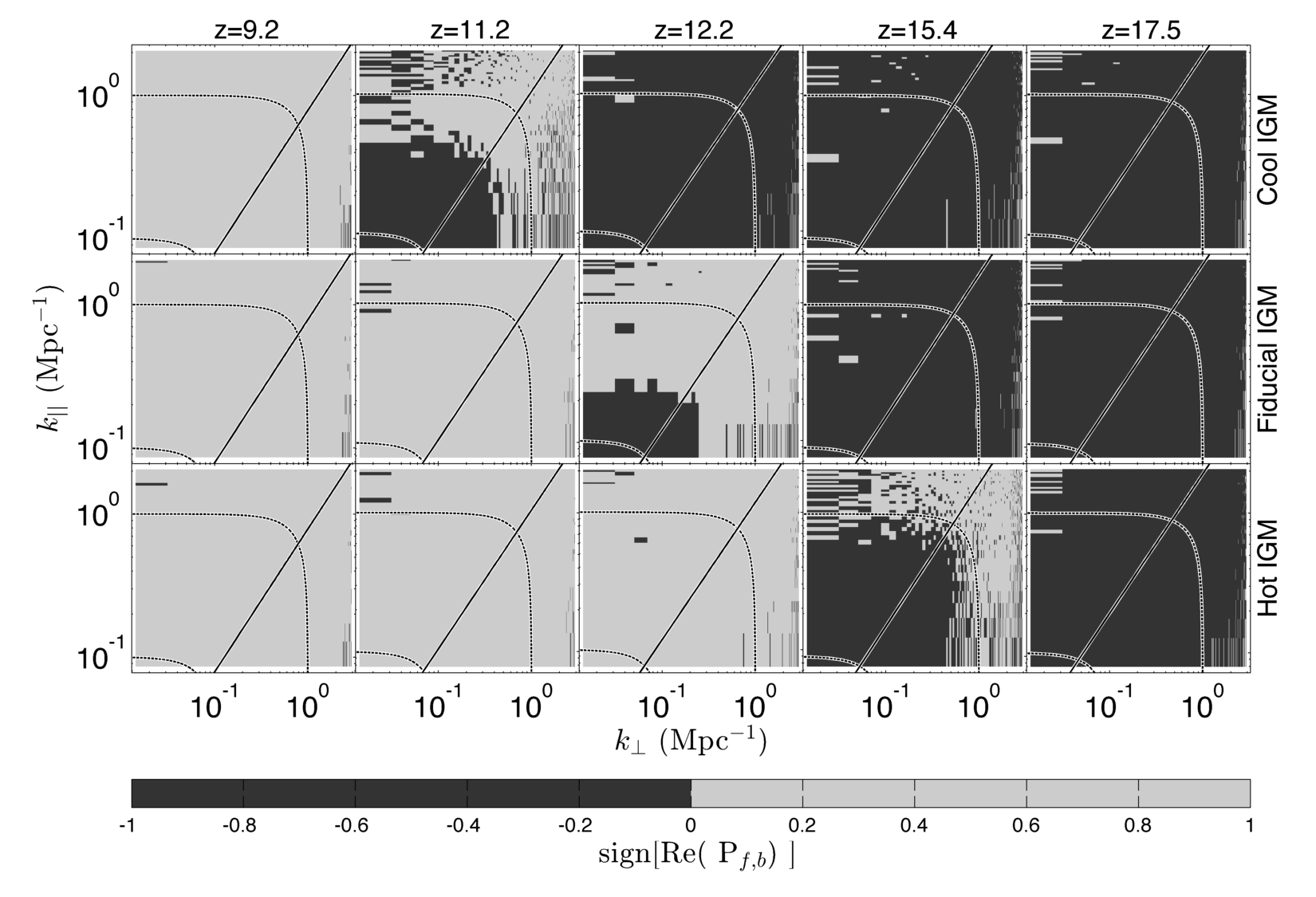}
\caption{The cross power spectrum, $\text{Re}(P_{f,b})$'s, sign is determined by the anti-correlation of $x_{HI}$ and $T_s$ during the pre-heating epoch and by $x_{HI}$ after heating has taken place. Here we show the sign of $\text{Re}(P_{f,b})$ for our three different heating models as a function of redshift. At pre-heating redshifts, $T_s$ is small and $x_{HI}$ is relatively uniform so that $\dtb$ and $T_f$ primarily depend on $T_s$ and anti-correlate so that $\text{Re}(P_{f,b})$ is negative. At low redshifts, $\dtb$ is independent of $T_s$ and fluctuations are primarily sourced by $x_{HI}$ so that $\dtb$ and $T_f$ are correlated and $\text{Re}(P_{f,b})$ is positive. Futhermore, heating proceeds in an ``inside-out" manner so that the smallest scales become correlated first.}
\label{fig:ps_modified_sign}
\end{figure*}

Verifying our prediction on the sign of $P_{f,b}$ is our next task; we plot this quantity in Figure \ref{fig:ps_modified_sign} for all models and redshifts. At high redshift, $P_{f,b}$ is entirely negative due to the anti-correlation between $T_f$ and $\dtb$ and adds to the total amplitude of $P_b'$. As heating takes place, $T_s$ drops out of $\dtb$ and fluctuations in $\dtb$ are sourced predominantly by variations in $x_{HI}$ leading to positive correlation between $\dtb$ and $T_f$ for positive $P_{f,b}$.  As we see from the figures, this process is ``inside-out'', with large scales remaining anti-correlated longer than the small scales. Heating proceeds in an ``inside-out'' manner, and since there is an overlap between the completion of heating and onset of reionization, temperature fluctuations remain important on large scales \citep{2007MNRAS.376.1680P,Mesinger:2013vm}.

\section{ PROSPECTS FOR DETECTION WITH AN MWA-LIKE ARRAY} \label{sec:det}

We now turn to addressing the detectability of the power spectrum signature of the forest and its distinguishability from the power spectrum, $P_b$. Our strategy is to combine our simulations with random realizations of instrumental noise and galactic and extragalactic foregrounds.  With data cubes containing both our simulated signals and our random contaminants, we can then take advantage of the full quadratic estimator formalism developed by \citet{Tegmark:1997wm}, adapted for 21 cm tomography by \citet{Liu:2011tt}; hereafter LT11, and accelerated for large data sets by \citet{Dillon:2013te};hereafter D13.  In this section, we will explain those techniques and show what results when our simulations of the forest are added to realistic foregrounds and instrumental noise.

\subsection{Power Spectrum Estimation Methods}

To estimate the power spectrum of the forest, we apply the quadratic estimator formalism \citep{Tegmark:1997wm}.  This formalism has the advantage that, in the approximation of foregrounds and noise that are completely described by their covariances, all cosmological information is preserved in going from three-dimensional data cubes to power spectra.  This formalism was adapted by LT11 for 21 cm power spectrum estimation and further refined and accelerated by D13.\footnote{For further details on this particular implementation of the quadratic estimator method, the reader is referred to D13.}  

In essence, the method relies an optimal and unbiased estimator of band powers in the $k_\perp$-$k_\|$ plane, $\widehat{\mathbf{p}}$, defined as
\be
\widehat{p}^\alpha = \sum_\beta M^{\alpha\beta} \left(\mathbf{x}^\trans\mathbf{C}^{-1}\mathbf{Q}^\beta\mathbf{C}^{-1}\mathbf{x} - b^\beta\right).
\ee
where $\mathbf{x}$ is a vector containing mean-subtracted data, $\mathbf{C}$ is the covariance of $\mathbf{x}$, including noise and contaminants, $\mathbf{Q}$ is a matrix that encodes the Fourier transforming, squaring, and binning necessary to calculate a band power, and $\mathbf{b}$ is the bias term.  The normalization matrix $\mathbf{M}$ is related to the Fisher information matrix $\mathbf{F}$.  Both $\mathbf{F}$ and $\mathbf{b}$ can be calculated via a Monte Carlo using the fact that
\be
b^\beta = \langle \mathbf{x}^\trans\mathbf{C}^{-1}\mathbf{Q}^\beta\mathbf{C}^{-1}\mathbf{x} \rangle  \equiv \langle \widehat{q}^\beta \rangle
\ee
and that
\be
\mathbf{F} = \text{Cov}(\widehat{\mathbf{q}}).
\ee

The ensemble average of each band power is related to the true band power $\mathbf{p}$ by a window function matrix, $\mathbf{W} = \mathbf{MF}$,
\be
\langle \widehat{\mathbf{p}} \rangle = \mathbf{Wp}.
\ee
The error on true band powers is also related to $\mathbf{M}$ and $\mathbf{F}$ through
\be \label{eq:cov}
\text{Cov}(\widehat{\mathbf{p}}) = \mathbf{MFM}^\trans.
\ee
Each quadratic estimator can thus be thought of as a weighted average of the true band powers with potentially correlated errors, both of which depend on one's choice of $\mathbf{M}$. Though any choice of $\mathbf{M}$ that makes $\mathbf{W}$ a properly normalized weighted average is reasonable, we adopt a form of $\mathbf{M}$ that makes the errors on $\widehat{\mathbf{p}}$ uncorrelated. \citet{Dillon:2014}, argue that this choice of $\mathbf{M}$ dramatically reduces the contamination of the EoR window by residual foregrounds.  It also provides a set of band power estimates which can be considered both mutually exclusive and collectively exhaustive because they cover the whole $k_\perp$-$k_\|$ plane while not containing any overlapping information.

\subsection{Noise and Foreground Models} \label{sec:CovModels}
The method outlined above requires model means and covariances of the contaminants that contribute to $\mathbf{x}$, like noise and foregrounds.  Our model of the instrumental noise depends, first and foremost, on the design of the interferometer.  In this paper, we consider the MWA with 128 tiles whose locations are detailed in \citet{Beardsley:2012ws} as representative of the current generation of low frequency interferometers.  Additionally, we consider possible realizations of double and quadruple sized instruments (MWA-256T and MWA-512T, respectively), as representative of extensions to current generation interferometers or next generation, $A_{eff} \sim 0.1 \km^2$, arrays such as the Hydrogen Epoch of Reionization Array (HERA) \citep{Backer:2009ve}.  As we will show, we generally do not need a square kilometer scale instrument to see the statistical effects of the forest.

\begin{figure}
\centering 
\includegraphics[width=.48\textwidth]{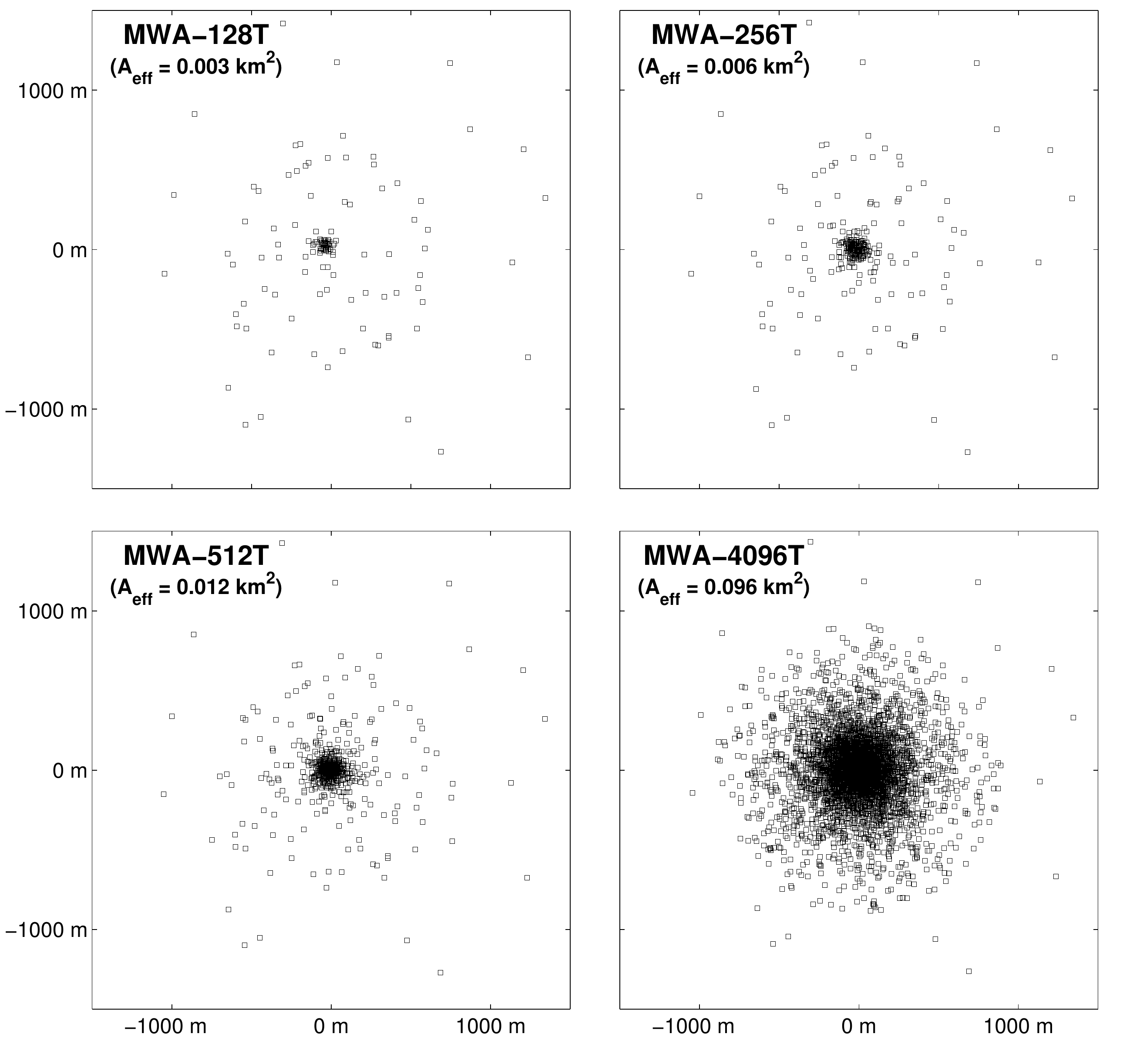}
\caption{Array layouts that we use to determine the detectibility and distinguishability of the 21 cm forest power spectrum signature. We chose to study two moderate extensions of the MWA-128T: MWA-256T and MWA-512T. In addition we study a 4096T array that is representative of a HERA scale instrument with $\sim 400$ times the collecting area of the MWA. Tile locations are drawn randomly from a distribution that is constant for the inner $50 \meter$ and drops as $r^{-2}$ for larger radii.}
\label{fig:ArrayLayouts}
\end{figure}

To generate our MWA-256T and MWA-512T designs with maximum sensitivity to 21 cm cosmology, we add antenna tiles to the current MWA-128T design within a dense core 900 m in radius.  These are drawn blindly from a probability distribution similar to that in \citet{Bowman:2007jw}: uniform for $r<50$ m and decreasing as $r^{-2}$ above 50 m.  The tile locations of the arrays we use are shown in Figure \ref{fig:ArrayLayouts}.

Our model for the noise is adapted from D13 \footnote{The method of D13 is adapted with one correction: the form of the noise power spectrum adapted from \citep{Tegmark:2009te} does not include the assumption that the field and beam sizes are the same.}.  In it, we incorporate observation times calculated in each $uv$-cell from 1000 hours of rotation synthesis at the latitude of the MWA.  The effective area of each tile is computed using a crossed dipole model while the system temperature is treated as the sum of receiver temperature, given by a power law fitted to two data points appearing in \citet{2013PASA...30....7T}, and sky temperature, measured in \citet{Rogers:2008vf}. In Table \ref{tab:mwa_params} we give our instrumental parameters at several different frequencies. 

\begin{table}
\centering
\caption{Instrumental Parameters}
\begin{tabular}{cccc} \hline\hline
f (MHz) &  FWHM (deg) & $A_{eff}$ (m$^2$) & $T_{sys}$ (K)\\ \hline
150 & 23 & 23& 290 \\ 
120 &30&24&  490\\ 
100	&34&24& 760\\ 
80  & 39&27&1300 \\ \hline
\end{tabular}
\label{tab:mwa_params}
\end{table}

Similarly, our model of the foregrounds is the one application of the model developed by LT11 and D13.  For the sake of simplicity,\footnote{Breaking extragalactic foregrounds into a bright ``resolved" population and a confusion-limited ``unresolved" population only improves the error bars (D13), so our efficient choice is also a conservative one.} we model extragalactic foregrounds as a random field of point sources with fluxes up to 200 Jy.  They have an average spectral index of 0.5 and variance in their spectral indices of 0.5.  Their clustering has a correlation length scale of $7'$.  Likewise, we model Galactic synchrotron radiation as a random field with an amplitude of 335.4 K at 150 MHz, a coherence length scale of $30^\circ$, and a mean spectral index of 0.8 with an uncertainty in that index of 0.1.

As we have previously discussed, we conservatively cut out the region of $k_\perp$-$k_\|$ space that lies below the wedge.  Once the wedge has been excised, we optimally bin from 2D to 1D Fourier space with the inverse covariance weighted technique described by D13.

To create simulated observations, we divide our simulated volumes into 36 fields, each 750 Mpc on a side, which roughly fill the primary beam of our antenna tiles.  We add random noise and foregrounds to each field independently, taking advantage of the fast technique for foreground and noise simulations developed by D13.  Finally, we take the sample variance of the cosmic signal into account by  using our power spectrum results from Section \ref{sec:results} and by counting the number of independent modes probed by the instrument at each $k$ scale.

\subsection{Detectability Results}
We now present the results of our sensitivity calculation. We demonstrate that, given prior knowledge\footnote{Here, ``prior knowledge" means that we know what the IGM power spectrum without the 21 cm forest  to within the error bars of our thermal senstivity.} of the X-ray heating history, a power spectrum measurement with a modest expansion of an MWA-like instrument is sufficient to distinguish between scenarios with or without the forest in our fiducial and cool heating models. Since the forest signal is detectable with smaller arrays only at smaller $k$, where $P_b$ dominates, its effect is likely degenerate with diffuse IGM emission. Observing this region for all considered models will require a HERA scale instrument with $A_{eff} \sim 0.1 \km^2$. 

\begin{figure*}
\centering 
\includegraphics[width=1\textwidth]{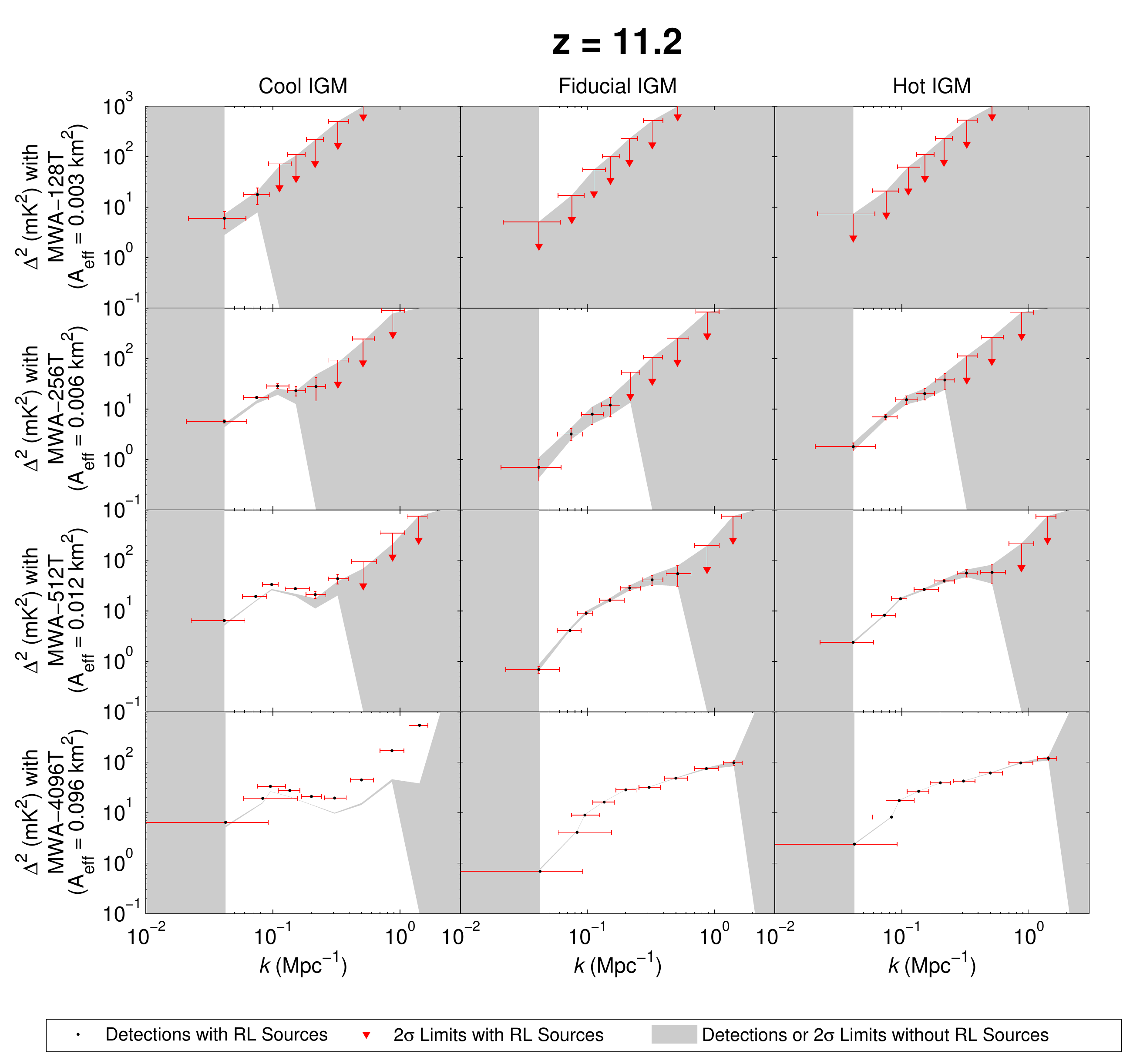}
\caption{Detections (black dots) and upper limits (red triangles) of the 21 cm Power Spectrum at z=11.2 for all of our arrays and heating models in the presence of 21 cm forest absorption from background RL sources. The grey fill denotes the 2 $\sigma$ region around the measured power spectrum with no RL sources present. To determine whether we can detect the forest imprint we ask, ``do the points and their error bars lie outside the gray shaded region?" MWA-256T and MWA-512T would be capable of distinguishing power spectra with or without sources in our cool IGM model, however only 4096T is consistantly sensitive to the $k \gtrsim 0.5 \Mpci$ region where the forest dominates. Only for our cool IGM model, MWA-512T would sufficient to detect this upturn as well. Hence a moderate MWA extension would likely be able to constrain some RL populations given a cooler heating scenario while a HERA scale instrument will be able to constrain the W08 RL population using the Forest power spectrum even for more emissive heating scenarios. Note that the upturn in the gray region is not from increased power at high $k$ but larger error bars.}
\label{fig:PlotModelVsArray}
\end{figure*}

In order to determine the array size necessary to resolve the forest power spectrum, we first focus on $z=11.2$, the lowest redshift considered where there is significant signal for one of our thermal models and quasar counts are relatively high. In Figure \ref{fig:PlotModelVsArray} we shade the $2 \sigma$ region for a detection of $\Delta^2(k)$ with no 21 cm forest absorption present and mark detections of $\Delta^2(k)$ with 21 cm forest absorption with black dots. The $2 \sigma$ vertical error bars, given by the diagonal elements of Equation (\ref{eq:cov}) are marked in red. Also marked in red are the horizontal error bars which are given by the 20$^{th}$ and $80^{th}$ percentiles of the window functions. To determine whether we can detect the forest imprint, we ask ``are the points consistent with the gray shaded region?"

We see that MWA-256T and MWA-512T can distinguish cool models with and without the forest at greater than $2 \sigma$. However these detections are not within the region of Fourier space where the forest dominates $P_b'$.  As a result, though MWA expansions can resolve two models with or without the forest, it is unlikely that they will be able to distinguish a model with the forest from one with a slight variation in heating. If an independent measure of the global spin temperature can be obtained, the radio luminosity function might be constrained with a modest MWA extension. We note that MWA-4096T is only able to detect the forest in our cool model at $z=11.2$ since the optical depth in our more X-ray emissive models is far too small at this time. 

\begin{figure*}
\centering 
\includegraphics[width=\textwidth]{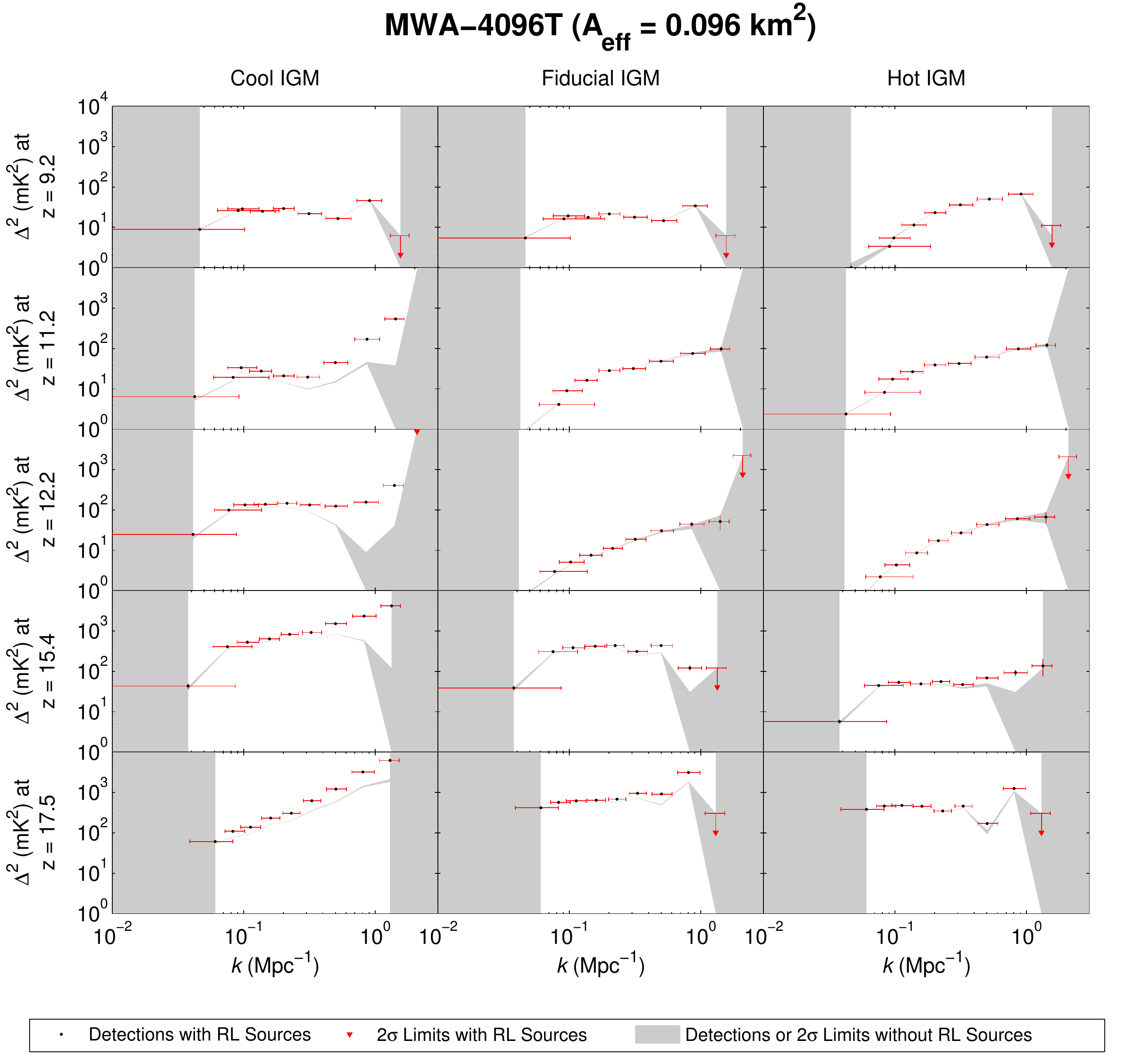}
\caption{These plots are identical to Figure \ref{fig:PlotModelVsArray} except the array is fixed to be MWA-4096T, representative of a HERA generation instrument, and redshift is varied. A HERA class instrument is able to resolve the upturn at $k \gtrsim 0.5 \Mpci$ that distinguishes the forest, and should be able to detect the 21 cm forest feature considered in this work for a variety of heating scenarios. The thermal noise error bars are to small to resolve by eye in most of these plots.}
\label{fig:PlotModelVsZ4096T}
\end{figure*}

To see more broadly what might be achieved by the next generation, we show in Figure \ref{fig:PlotModelVsZ4096T} the error bars and detections with and without the forest across all considered $f_X$ and $z$ for our HERA scale model. We find that z=15.4 is our ``sweet spot" for the W08 distribution. 4096T is able to resolve the $k \gtrsim 0.5 \Mpci$ forest region for all of the IGM heating models that we investigate. For our cool and fiducial models, 4096T is also able to observe the forest region for a range of redshifts. These results show that a HERA scale array has the potential to constrain the IGM state by measuring $\Delta^2$ for $k \lesssim 10^{-1} \Mpci$, where the brightness temperature dominates, and the RL distribution in observing the region $k \gtrsim 0.5 \Mpci$ where the forest has a significant contribution.

Over the course of the IGM's evolution, there are times where the 21-cm power spectrum becomes particularly steep; for example, during the era immediately before the X-ray heating peak.  As a result, observing excess power at $k \gtrsim 0.5 \Mpci$ for a single redshift alone will likely not be sufficient to constrain the radio luminosity function. However, discerning the IGM thermal history with measurements of the power spectrum amplitude at $k \sim 0.1 \Mpci$ and observing an absence of flattening at high k, over the range of redshifts after the X-ray heating peak as shown in Figure \ref{fig:ps} should allow for constraints to be placed on the high-redshift radio luminosity function.

\subsection{Distinguishability Results}

In order to quantify how distinguishable our simulations with the forest are from our simulations without the forest for a given instrument, redshift, and heating model, we calculate the standard score of the $\chi^2$ sum of the power spectrum values across all k-bins,  
\begin{equation}\label{eq:zscore}
Z \equiv  \frac{\chi^2-N_k}{\sqrt{2N_k}},
\end{equation}
where  $N_k$ is the number of k bins, $\chi^2 \equiv \sum_k \left(\frac{P_b(k)'-P_b(k)}{\sigma_k} \right)^2$, and $\sigma_k$ are the diagonal elements of Equation (\ref{eq:cov}) for each model without the 21 cm forest present. Assuming statistical independence between $k$ bins, $Z$ is the number of standard deviations at which we can distinguish a model with the 21 cm forest from a model without it using the $\chi^2$ statistic. Unfortunately, this measure is somewhat naive since it does not account for potential degeneracies in the power spectrum amplitude from different thermal histories. However it enables us to quantitively compare outlooks across the numerous dimensions of redshift, array, and heating history. We consider a $Z \gtrsim 10$ to indicate significant distinguishability.  

\begin{figure*}
\centering 
\includegraphics[width=\textwidth]{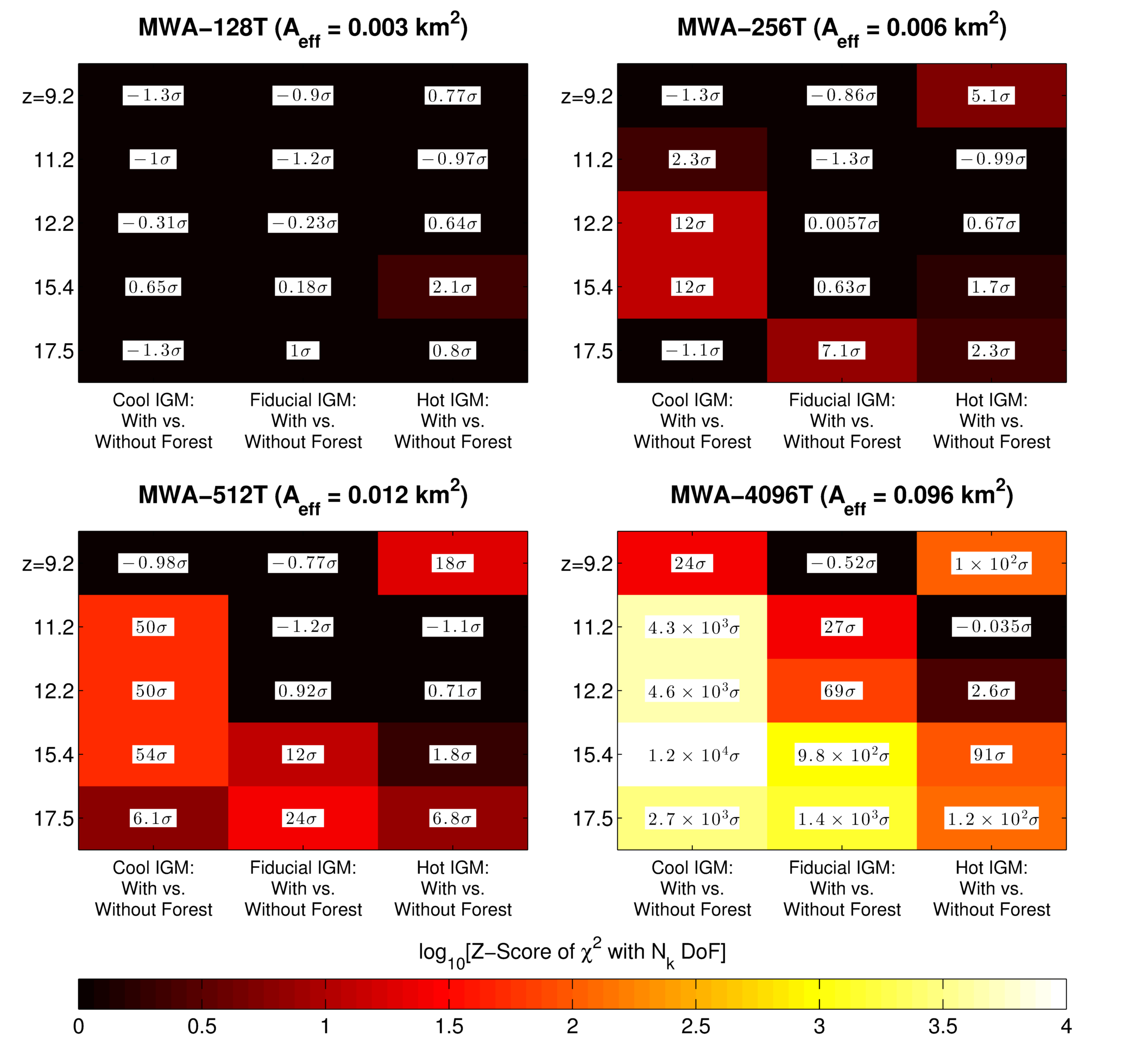}
\caption{The significance of distinguishability across all measured k bins (Equation (\ref{eq:zscore})) for all arrays, redshifts, and IGM heating models for a 1000 hour observation. An extension of MWA-128T is capable of distinguishing models with and without the 21 cm forest from the W08 RL population in our cool and fiducial heating scenarios. MWA-512T and HERA scale MWA-4096T are capable of distinguishing the forest in the power spectrum in all heating models considered in this work.}
\label{fig:PlotModelZScore}
\end{figure*}

In Figure \ref{fig:PlotModelZScore} we show  the value of Equation (\ref{eq:zscore}) for all models and arrays. Our first observation is that MWA-128T is not capable of distinguishing a model with the forest from a model without the forest for any of the considered $f_X$. MWA-256T would be capable of distinguishing the forest at all considered $z\gtrsim 9.2$ for our cool X-ray heating model at greater than $5 \sigma$ and in our fiducial heating model only at the highest considered redshift (which is near the X-ray heating peak). MWA-512T would be capable of resolving the forest at the two highest redshifts for our fiducial model and at all considered redshifts for our cool model. The hot model remains unobservable for all MWA expansion arrays but is accessible to a HERA scale instrument.

\begin{figure*}
\centering
\includegraphics[width=\textwidth]{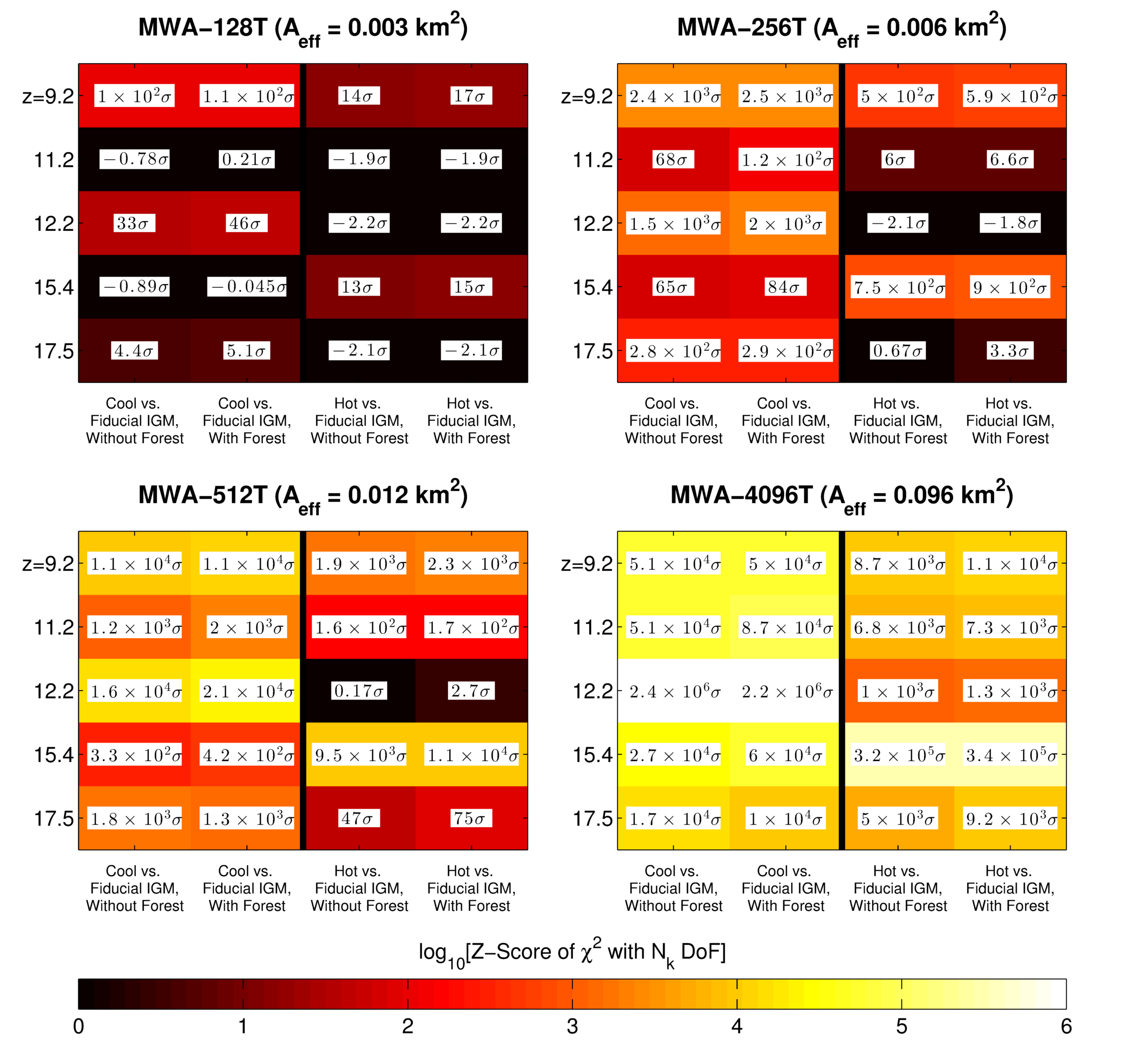}
\caption{ The 21 cm Forest moderately enchances the distinguishability between thermal scenarios and MWA scale interferometers can distinguish between the power spectra for reasonable X-ray heating histories. Here we show the cumulative z-score described in Equation (\ref{eq:zscore}), except now applied to the difference between different IGM heating models, for all arrays and redshifts. At low redshift, the forest decreases the distinguishability of different X-ray heating scenarios by subtracting from the higher amplitude model. When the positive auto-term dominates at high redshift, the forest increases the contrast between given heating models.}
\label{fig:PlotIGMZScore}
\end{figure*}

How the distinguishability between different heating models is affected by the presence of the 21 cm forest is explored in Figure \ref{fig:PlotIGMZScore}. In our 128T table, we see that a  detection of the IGM and constraints on low X-ray emissive histories are possible with the current generation of EoR experiments. There are several caveats worth noting however. First, the high S/N distinctions at $z=9.2$ are due to a detection of the reionization peak at redshifts in which reionization physics such as the uv-efficiency (which we have assumed fixed) become significant.  However, we note that this result contradicts the marginal detectability claimed in \citet{Mesinger:2013c} primarily due to the fact that we include bins with $k < 0.1 \Mpci$ in our standard score. Though these bins have large S/N they may be contaminated by more pessimistic foreground leakage than we consider here such as what is observed by \citet{Pober:2013tg}.  We also note that the increased sensitivity of combining k-bins allow for constraints on the fiducial X-ray model at $z \sim 15$. The peaks in detectability at $z \approx 9$ and $z \approx 17$ arise from the two peaked structure of the power spectrum in redshift with the low redshift peak corresponding the reionization, and the high redshift peak corresponding to x-ray heating \citep{2007MNRAS.376.1680P}.  We see that the forest introduces a small enhancement to the distinguishability between hot and cool heating models. Since the forest adds positively to the power spectrum of a cool, optically thick IGM, its presence enhances the distinguishability between vigorous and cool heating. We find that a modest extension to the MWA can distinguish between hot and fiducial models over a wider range of redshifts and MWA-4096T is able to distinguish between all models over our entire considered redshift range. 

\section{The Detectability of the Forest over a Broad Parameter Space}\label{sec:extr}

\begin{figure*}
\includegraphics[width=\textwidth]{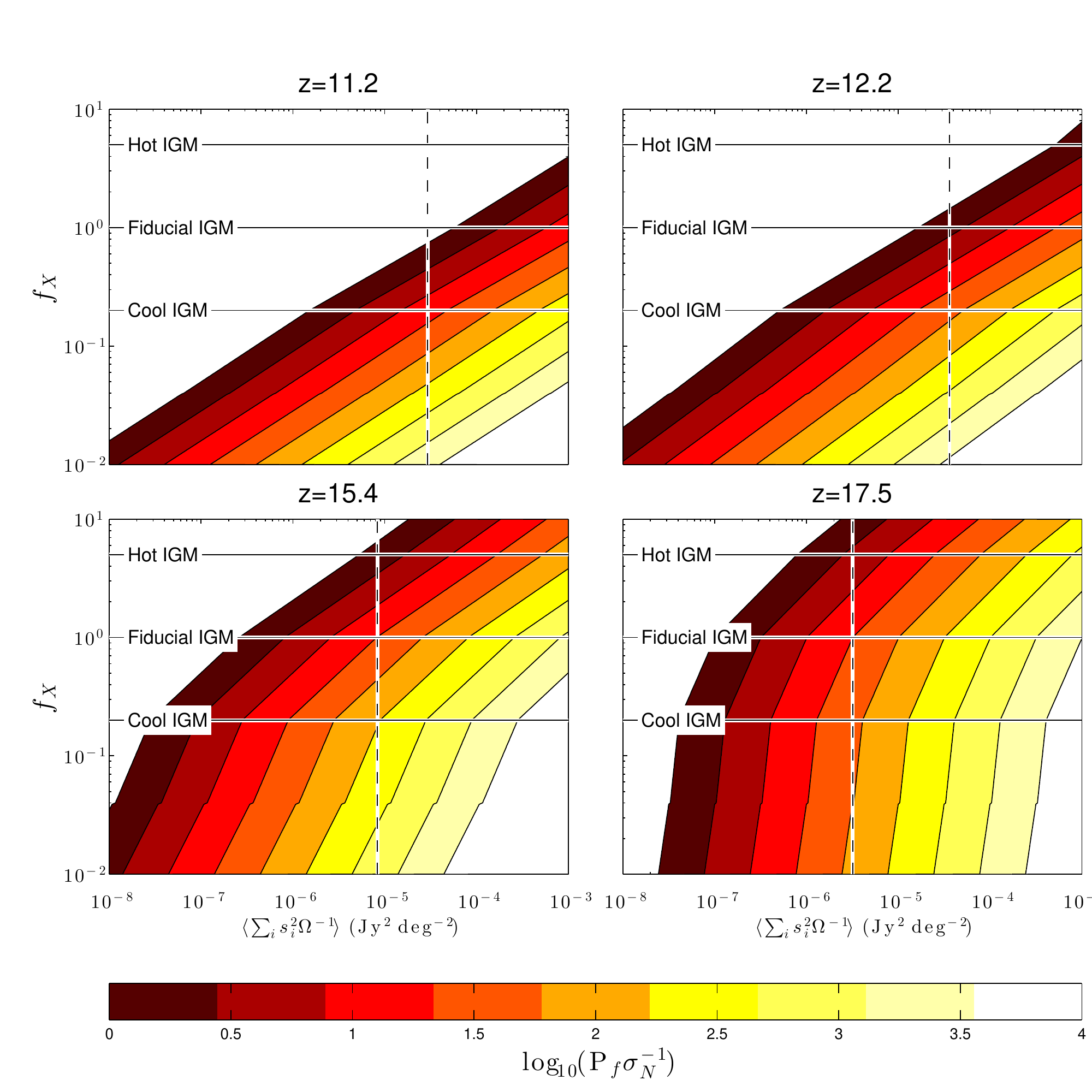}
\caption{The ratio of the 21 cm Forest power spectrum, $P_f(k=0.5 \Mpci)$ to thermal noise for 1000 hours of observation on a HERA scale interometer, extrapolated over a large range of X-ray efficiencies and flux squared densities. Vertical dashed black and white lines indicate the value of the simulation by \citep{2008MNRAS.388.1335W} while the horizontal black and white lines indicate the $f_X$ efficiencies that we explicitly simulate in this paper. At the highest redshifts, $\langle T_s \rangle$ levels off and the detectability of the signal is independent of redshift. At late prereionizatoin redshifts, we see that the 21 cm Forest will only be detectable for heating efficiencies $\lesssim 1$.}
\label{fig:sn_params}
\end{figure*}

For the sake of simplicity, we focus on the detectability of the 21 cm Forest power spectrum from the single population model considered in \citet{2008MNRAS.388.1335W}. In doing this, it is unclear over what range of radio loud populations the signal is observable. Fortunately, thanks to  Equation (\ref{eq:integral_fix}), we can give order of magnitude estimates of how the detectability of the Forest power spectrum scales with the radio loud source population and the heating history. According to Equation (\ref{eq:integral_fix}), the amplitude of the forest power spectrum, at prereioinization redshifts, scales as 
\begin{equation}\label{eq:scaling}
P_f \propto \frac{1}{\langle T_s \rangle^{2}} \frac{ \sum_i s^2_i(>z) }{\Omega } 
\end{equation}
 where $\sum_i s^2_i(>z) \Omega^{-1}$ is the average sum of source fluxes squared per solid angle. We will call this quantity the {\it flux squared density} of the source population. We take advantage of the simple scaling  in Equation (\ref{eq:scaling}) to extrapolate the amplitude of the Forest signal over a large range of heating models and redshifts. At each redshift, with our fiducial heating model and source population, we obtain a normalization factor for  $P_f$ at a single mode, $k = 0.5 \Mpci$. We then compute $\langle T_s \rangle $ for a large number of lower resolution, $(600 \Mpc)^3$ 21cmFAST simulations with $400^3$ pixels, varying the $f_X$ parameter by three orders of magnitude from $f_X = 10^{-2} - 10^1$. In Figure \ref{fig:sn_params}, we show the ratio of $P_f$ to the amplitude of thermal noise as a function of $f_X$ and the flux squared density of sources, marking the predicted flux squared density of \citet{2008MNRAS.388.1335W} by a dashed black and white line. We find that the detectability of the forest power spectrum at $z \sim 10$ depends strongly on the thermal state of the IGM, with models significantly fainter than \citet{2008MNRAS.388.1335W} undetectable except for cool heating histories with  $f_X \lesssim 10^{-1}$. On the other hand, for $z \gtrsim 15$, X-rays in all models have not had sufficient time to heat the IGM above the adiabatic cooling floor and the detectability of $P_f$ becomes significantly less dependent on $f_X$, allowing for a broader range of populations to be probed at higher $f_X$.

\section{Conclusions and Future Outlook}

Using semi-numerical simulations of the thermal history of the IGM, and a semi-empirical RL source distribution, we have shown that the 21 cm forest imprints a distinctive feature in the power spectrum that is, for the most part, invariant in $\kperp$ and, depending on the RL population and thermal history, potentially dominates over the cosmological 21 cm power spectrum at $k_\parallel \gtrsim 0.5 \Mpci$.  We have also derived a simple semi-analytic equation that directly relates the forest power spectrum of $\tau_{21}$ and the radio luminosity function. 
 
Using realistic simulations of power spectrum estimation and including the effects of foregrounds and noise, we have shown that a moderate extension of the MWA-128T instrument has the thermal sensitivity to detect the forest feature in the power spectrum for the W08 RL population with an X-ray efficiency of $f_x \lesssim 1$. For more vigorous heating scenarios, a HERA scale array will have the sensitivity to distinguish this feature. Our simulations also support the results of \citet{Christian:2013} and \citet{Mesinger:2013c}, that low emissivity heating scenarios can be constrained with existing arrays and an extensive examination of the heating history will be possible in the future with larger instruments.    

Signal-to-noise considerations alone do not tell us whether we will be able to distinguish the forest signal from the effects of IGM physics on the power spectrum, especially at small $k$ where a slight change in $f_X$ might shift the power spectrum amplitude up or down, mimicking the shift from the 21 cm forest. Fortunately, the region, $k \gtrsim 0.5 \Mpci$ is dominated by the forest power spectrum, $P_f$, for a range of redshifts in all of our heating models. Specifically, the 21 cm forest removes the $k \gtrsim 0.5 \Mpci$ flattening that occurs after the X-ray heating peak. Observations of the power spectrum over a range of redshifts, with a sensitivity similar to HERA or the SKA should be able to isolate the thermal history at $k \lesssim 0.1 \Mpci$ and constrain RL populations similar to that of W08 at $k \gtrsim 0.5 \Mpci$.

While this paper is a proof of concept, considering a single fiducial RL source distribution, it is possible that measurements with current generation instruments, or moderate extensions, can put constraints on more optimistic scenarios. On the other hand, there are many steep decline scenarios whose power spectrum signatures will be inaccessible even to future arrays. In section \ref{sec:extr} we illustrate the scaling of the detectability of the signal with source flux squared density and X-ray emissivity, finding that populations with order of magnitude smaller flux squared densities than W08 will require a relatively cool prereionization IGM  to be detectable.  In particular, we note that the H04 simulation is one to two orders of magnitude more pessimistic than the predictions of W08 at the highest considered redshifts and would not be detectable in the forest dominated region if $f_X \gtrsim 10^{-1}$. However, higher resolution simulations of the IGM indicate that $\Delta^2_{\tau_{21}}$ continues to climb to $k \sim 10 \Mpci$ while $P_b$ remains flat. Hence the result of a fainter radio luminosity function would be to shift the region of forest dominance to higher k rather than eliminating it, leaving the possibility of detection for a more powerful instrument such as the SKA. There also exists the possibility of separating $P_f$ using its LoS symmetry which might be exploited at $k \sim 0.1 \Mpci$ where EoR interferometers are most sensitive. Finally, we have not considered the absorption of mini halos which \citet{Mack:2012tf} show to substantially increase the variance along the line of sight towards sources (see their Figure 11). Since this variance is an integral of the power spectrum we are being conservative in neglecting them. The sensitivity of future instruments to the forest can be enhanced by increased frequency resolution, allowing them to probe the higher $\kpara$ modes where the forest is especially strong.  The parameter space of radio loud quasars is greatly unconstrained and the disparity between W08 and H04 simply underscores the need for future studies to explore this parameter space. The exploration of a range of RL populations for fixed arrays is left for future work.

In summary, we have shown that the 21 cm power spectrum not only contains information on the IGM in absorption and emission against the CMB but also includes detectible, and in many cases non-negligible signatures of the 21 cm forest. This absorption may be used to constrain the high redshift RL population and IGM thermal history with upcoming interferometers.

\section{Acknowledgements}
The authors would like to thank A. Liu, M. Matejek, and M. Tegmark for some informative discussions. A. E-W., J. S. D., and J. H. thank the Massachusetts Institute of Technology School of Science for support. A. E-W. acknowledges support from the National Science Foundation Graduate Research Fellowship Grant No. 1122374 and the Bruno Rossi Fellowship.

\appendix
\section{A Derivation of the morphology of $P_f$}\label{app:compare_pf}
In Section \ref{sec:theory} we present a formula, Equation (\ref{eq:pf_simple}), for the the 21 cm forest power spectrum that is the sum of the auto power spectra along the line of sight to each background source. This equation is particularly convenient because it can easily be decomposed into an integral of the radio luminosity function and the optical depth power spectrum. In addition, its $k$-space morphology, which includes no structure in $\kperp$, is relatively simple. 
 In this appendix we derive Equation (\ref{eq:pf_simple}) by applying an analytic toy model to the auto and cross power spectrum contributions to $P_f$ described in Equation (\ref{eq:ps_with_RL}). For the sake of analytic tractability, we invoke a number of approximations. However our results describe $P_f$ very well for $k \gtrsim 10^{-1} \Mpci$. Our assumptions are
\begin{enumerate}
\item  The sources all have the same flux. The W08 simulation includes sources ranging from $1 \nJy$ to $\sim10 \mJy$ over the redshifts of interest. We see in Figure \ref{fig:flux_integral} that the integral of the source fluxes squared is dominated (at the 10\% level) by sources with $S_\nu$ between $1-10\mJy$ so modeling our population as having equal flux gives a decent order of magnitude approximation.
\item The sources are spatially uncorrelated. Clustering from the W08 dark matter bias is actually significant and boosts the results of our simulation, relative to Equation (\ref{eq:pf_simple}), by a factor of two without changing $P_f$'s predicted shape. We will thus absorb this clustering boost into a multiplicative factor of order unity.
\item The sources are unresolved. This will almost certainly be true in all interesting cases given the large synthesized beams of radio interferometers and the extreme distances to the sources. 
\item Source spectra are flat over the frequency interval of a data cube. This is true on the 10\% level over a $\sim 8 \MHz$ band for $S \sim \nu^{-.75}$ sources. Because this slow variation gives a very narrowly peaked convolution kernel in $k$-space, power spectra are not noticably effected by this assumption.
\item The source positions are completely uncorrelated with the cube optical depth field. In reality, the sources that fall within a data cube should be correlated with $\tau_{21}$. We find that correlating or not correlating in cube sources only changes the simulation output by approximately $10\%$.
\end{enumerate}

  We start by reiterating Equation (\ref{eq:sum_forest_sources}) where $P_f$ may be written as
\begin{align} \label{eq:sum_forest_app}
P_f  &= \frac{1}{V} \left \langle \left| \widetilde{\Delta T_{RL} \tau_{21}  }\right|^2  \right \rangle \notag \\
& = \displaystyle \sum_{j}   P_j +2   \text{Re} \left( \displaystyle \sum_{j < k}  P_{j,k} \right) \notag \\ 
& \equiv \Sigma_{auto}+\Sigma_{cross},
\end{align}
where $P_j = \frac{1}{V} \langle | \widetilde{ \Delta T_j \tau_{21}}|^2 \rangle$ and $P_{j,k} = \frac{1}{V}\langle  \widetilde{ \Delta T_j \tau_{21}} \widetilde{ \Delta T_k \tau_{21} }^* \rangle$.
The first term in Equation (\ref{eq:sum_forest_app}) sums the power spectra of each of the absorbed background sources which is positive and the second term is the sum of their cross power spectra. 

  We will show that for the range of spatial scales perpendicular to the LoS, accessed by EoR interferometers, the auto power terms in Equation (\ref{eq:sum_forest_app}) dominate the cross power ones at $\kpara \gtrsim 10^{-1} \Mpci$. We show that the suppression of cross terms is due to two mechanisms: (1) the cross terms are proportional to the cross power spectra between widely separated lines of sight and (2) the cross terms are multiplied by randomly phased sinusoids which cancel out when summed.

\subsection{The Suppression of the Cross terms from LoS Cross Power Spectra}

 To relate the sum in Equation (\ref{eq:sum_forest_app}) to the spectra and locations of the background sources, we assume that all sources are unresolved so that $T_j$ is a delta-function in the plane perpendicular to the LoS. Here, as in \citet{McQuinn:2006ty}, we will adopt observers coordinates $(\ell,m,\nu)$, rather than comoving coordinates $(x,y,z)$, to emphasize the fact that the the broad-spectrum source does not physically occupy a range of positions along the LoS. In such coordinates, the temperature field of each source can be written as $T_j( \ell, m, \nu)$ where $\ell$ and $m$ are the direction cosines from the north-south and east-west directions, and $\nu$ is the difference from the data cube's central frequency. $\tau_{21} T_j(\ell, m,\nu)$ is given by
\begin{equation}\label{eq:single_source}
\tau_{21} T_{j}(\ell,m,\nu)=   \Omega_{pix} \delta(\ell - \ell_j) \delta( m - m_j) \tau_{21}(\ell_j, m_j, \nu)T_j
\end{equation}
where $ \Omega_{pix}$ is the solid angle of a map pixel and $\delta( ... )$ is the Dirac delta function. For notational simplicity, we will use vector notation to denote direction cosines, ${\bf \ell}=(\ell,m)$ and their Fourier duals, ${\bf u} = (u,v)$. Taking the Fourier transform of  $\tau_{21} T_j({\bf \ell}, \nu)$ and summing over all sources gets
\begin{align} \label{eq:inter}
\widetilde{T}_f({\bf u},\eta)  = &\sum_j \widetilde{\tau_{21}T_j}({\bf u},\eta) \notag \\
  =   \Omega_{pix}  &\sum_j T_j e^{2 \pi i ( {\bf \ell}_j \cdot {\bf u})}  \int \tau_{21}({\bf \ell}_j,\nu)  e^{-2 \pi i \eta \nu}  d \nu.
\end{align}
We take the modulus squared of Equation (\ref{eq:inter}) and multiply by the cosmology dependent variables, $D_M^2Y$ \citep{Parsons:2012vk} that relate observers coordinates to the cosmological comoving coordinates that we've used to define our power spectrum in Equation (\ref{eq:ps}). We find that the sum of the auto terms in Equation (\ref{eq:sum_forest_app}) is
\begin{equation}
\Sigma_{auto}= \frac{ D_M^2 \Omega_{pix}^2}{\Omega_{cube}}  P_{\tau_{21}}^{LoS}(\kpara) \left \langle \sum_j T_j^2 \right \rangle.
\end{equation}
The sum of cross terms is
\begin{align}
 \Sigma_{cross} &= 2 \frac{ D_M^2 \Omega_{pix}^2}{\Omega_{cube}}   \sum_{j<k}T_j T_k \Bigl[  \notag \\
& \text{Re} \left( P_{\tau_{21;j,k}}^{LoS}(\kpara)\right)\langle \cos [ 2 \pi( {\bf u} \cdot  {\bf \Delta \ell_{j,k} })] \rangle \notag  \\
+& \text{Im} \left( P_{\tau_{21};j,k}^{LoS}(\kpara)\right) \langle \sin[ 2 \pi( {\bf u} \cdot {\bf \Delta \ell_{j,k} })] \rangle \Bigr] \notag \\
&= 2  P_{\tau_{21}}^{LoS}(\kpara) \frac{ D_M^2 \Omega_{pix}^2 }{\Omega_{cube}} \sum_{j<k}T_j T_k \Bigl[ \notag \\
&  \frac{\text{Re}\left(P_{\tau_{21};j,k}^{LoS} (\kpara)\right)}{P_{\tau_{21}}^{LoS}(\kpara)}  \langle \cos [ 2 \pi  ( {\bf u} \cdot {\bf \Delta \ell_{j,k} })]  \rangle +\notag  \\ 
  &\frac{\text{Im}\left(P_{\tau_{21};j,k}^{LoS}(\kpara)\right)}{P_{\tau_{21}}^{LoS}(\kpara)} \langle \sin[ 2 \pi (  {\bf u} \cdot {\bf \Delta \ell_{j,k} } ) ] \rangle  \Bigr],\label{eq:cross_sum}
\end{align}

where ${\bf \Delta \ell_{j,k}}={\bf  \ell_j} - {\bf \ell_k}$. Here, we define the cross power spectrum between two lines of sight to be 
 
\begin{equation}
P_{\tau_{21};j,k}^{LoS}(k_z)=\frac{1}{L} \int d z d z'  e^{  i k_z (z-z')} \Delta \tau_{21} ( {\bf \ell}_j,z) \Delta \tau_{21}( {\bf \ell}_k  ,z').
\end{equation}

\begin{figure}
\centering
\includegraphics[width=.48\textwidth]{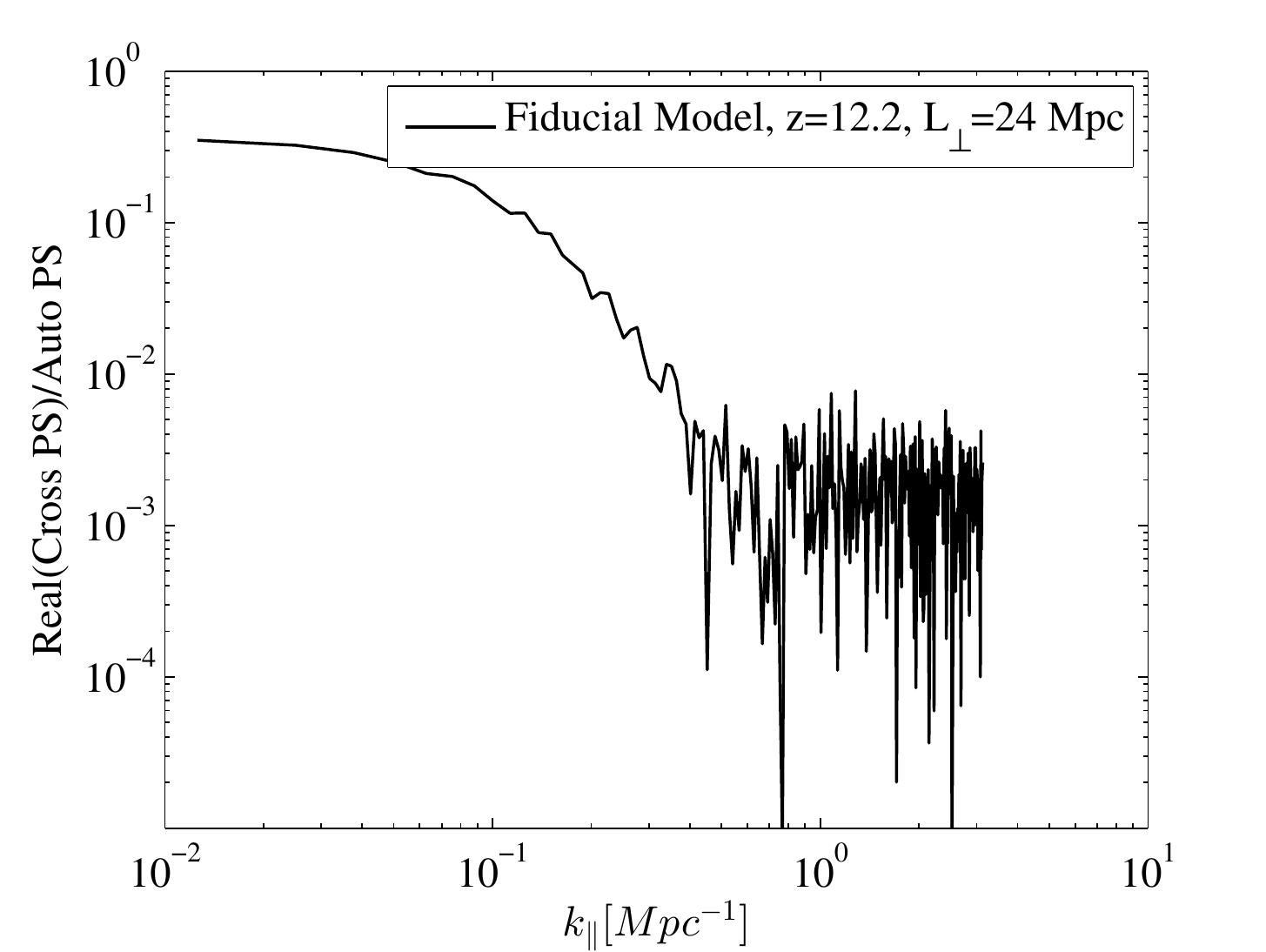}
\includegraphics[width=.48\textwidth]{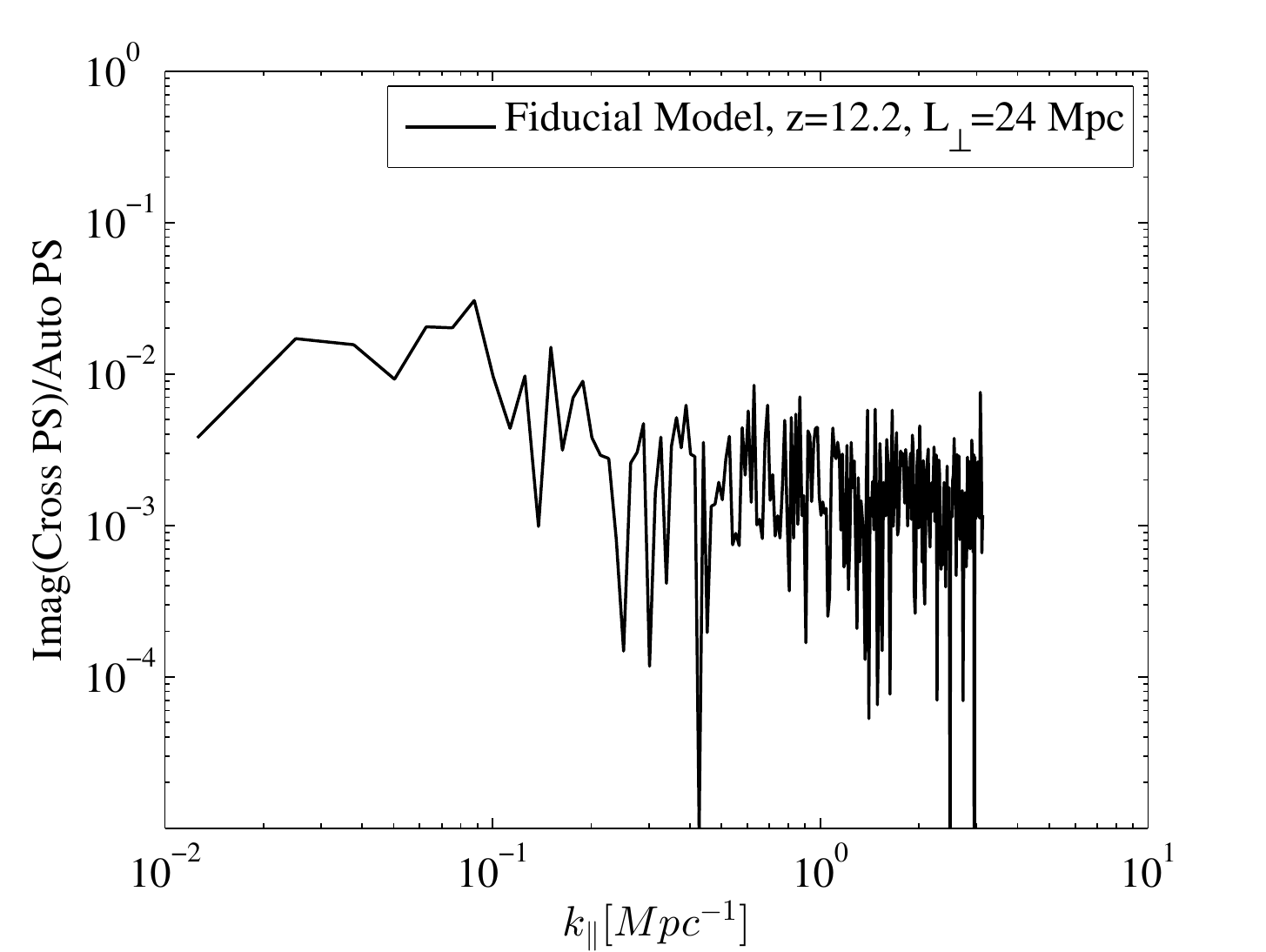}
\caption{The LoS cross power spectra between spatially separated lines of site are on the order of $\sim 100-1000$ times smaller than auto power spectra. In the left figure, we plot the ratio of the real cross power spectra between lines of site separated by 24 Mpc to auto power spectra, and on the right we show the ratio of the imaginary cross power spectrum to the auto power spectrum. In both cases, for $\kpara \gtrsim 10^{-1} \Mpci$, the cross power spectra are on the order of 10-1000 times smaller. The real cross power spectrum becomes non negligible on scales comparable to the separation between the two lines of site.}
\label{fig:cl}
\end{figure}

 It is clear from Equation (\ref{eq:cross_sum}) that each summand in $\Sigma_{cross}$ is smaller than each term in $\Sigma_{auto}$ by a factor of the ratio between the LoS cross power spectra of spatially separated lines of sight and the LoS auto power spectrum. If lines of sight to each source are sufficiently separated, this ratio should be very small. In Figure \ref{fig:cl} we show the ratios of $\text{Re} \left( P_{\tau_{21};j,k}^{LoS} \right)/P_{\tau_{21}}^{LoS}$ and $\text{Im} \left( P_{\tau_{21};j,k}^{LoS} \right)/P_{\tau_{21}}^{LoS}$ from our fiducial model at $z=12.2$, separated by $L_\perp=24 \Mpc$ which is the mean distance in our data cube between 1000 background sources. Because two sufficiently separated lines of sight should be statistically independent except on the largest spatial scales, these ratios are on the order of $10^{-2}-10^{-3} $ for $\kpara \gtrsim 10^{-1} \Mpci$ .

\subsection{Supression of the Cross Terms from Summing the Random Source Phases}

 The factor of 100-1000 introduced by the ratio of the cross spectra to the auto spectra would be enough to suppress the cross terms if the number of sources were reasonably small. However the number of cross terms relative to the number of auto terms in Equation (\ref{eq:sum_forest_app}) goes as $(N-1)/2$ where $N$ is the number of contributing sources. Thus, even though the cross power spectrum between individual LoS pairs is small, naively summing 100-500 sources could still yield a significant contribution. We now show that summing over many randomly distributed source angles suppresses this. 
 
\begin{figure}
\centering
\includegraphics[width=.48\textwidth]{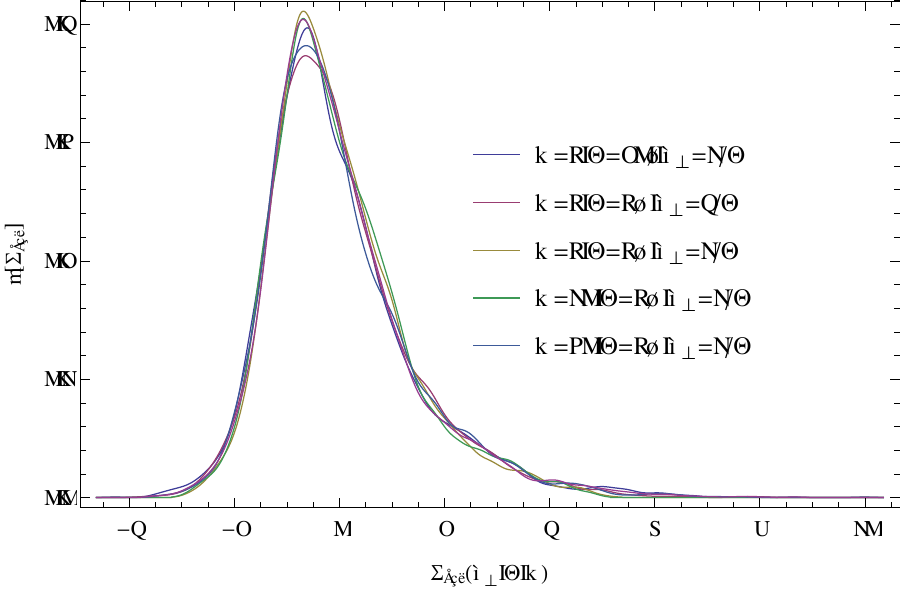}
\caption{Here we see that $P[\Sigma_{cross}(u_\perp,\Theta,N)$ is invariant in N, $\Theta$, and $u_\perp$, and $N$ with a mean of approximately zero. The lines which indicate, $P[\Sigma_{cross}(u_\perp,\Theta,N)]$, are estimated from 10000 draws. Since $\langle \Sigma_{cross} \rangle \approx 0$ we expect the cross terms to contribute negligibly to $P_f$ in 3D Fourier Space.}
\label{fig:dist_upper}
\end{figure}

Since $\text{Im} \left( P_{\tau_{21};j,k}^{LoS} \right)/P_{\tau_{21}}^{LoS}$ is on the same order of, or smaller than  $\text{Re} \left( P_{\tau_{21};j,k}^{LoS} \right)/P_{\tau_{21}}^{LoS}$, we will use the real term on both the cosine and sine terms in Equation (\ref{eq:cross_sum}) to give an upper bound. Assuming that all sources have the same temperature, $T_j=T_k=T_0$, we may write
\begin{align}\label{eq:sum_cross_full}
\Sigma_{cross} \approx  2 \sum_{j<k}T_0^2 P_{\tau_{21};j,k}^{LoS} (\kpara) \Bigl[ &   \langle \cos [ 2 \pi (  {\bf u} \cdot {\bf \Delta \ell_{j,k} })] \rangle \notag \\
 + &  \langle \sin [ 2 \pi(  {\bf u} \cdot {\bf \Delta \ell_{j,k} } )] \rangle \Bigr].
\end{align}
Similarly,
\begin{equation}
\Sigma_{auto} \approx N T_0^2 P_{\tau_{21}}^{LoS}
\end{equation}
Hence the ratio between $\Sigma_{cross}$ and $\Sigma_{auto}$ is given by
\begin{align}
\Sigma_{cross} / \Sigma_{auto} \approx  2 \frac{ \text{Re}\left( P_{\tau_{21};j,k}^{LoS}(\kpara) \right)}{N P_{\tau_{21}}^{LoS}(\kpara)}\sum_{j<k} \Bigl[& \langle \cos [ 2 \pi (  {\bf u} \cdot {\bf \Delta \ell_{j,k} })] \rangle\notag  \\
+ & \langle \sin [ 2 \pi (  {\bf u} \cdot {\bf \Delta \ell_{j,k} })]  \rangle \Bigr]
\end{align}
Because of the cylindrical symmetry, we need only concern ourselves with a uv cell at $v=0$ and simply write
\begin{align}
\Sigma_{cross} / \Sigma_{auto} \approx 2  \frac{\text{Re}\left( P_{\tau_{21};j,k}^{LoS}(\kpara) \right)}{N P_{\tau_{21}}^{LoS}(\kpara)} \sum_{j<k}  \Bigl[  &\langle \cos [ 2 \pi  u_\perp \Delta \ell_{j,k}]  \rangle \notag \\
+ &\langle \sin [ 2 \pi  u_\perp \Delta \ell_{j,k}] \rangle \Bigr] \notag \\
=   \frac{\text{Re}\left( P_{\tau_{21};j,k}^{LoS}(\kpara) \right)}{ P_{\tau_{21}}^{LoS}(\kpara)} & \left \langle \Sigma_{cos} \left( u_\perp,\Theta,N\right) \right \rangle,
\end{align}
where 
\begin{equation}\label{eq:cossum}
\Sigma_{cos} \left( u_\perp,\Theta,N \right) \equiv \frac{2}{N} \sum_{j<k}  \cos[ 2 \pi u_\perp \Delta \ell_{j,k}]  +  \sin[ 2 \pi u_\perp \Delta \ell_{j,k}]. 
\end{equation}
We can easily compute this ensemble average for any $u_\perp$ by drawing N different source positions  distributed randomly over the angular span of the field, $\Theta$, and summing over the sines and cosines of pair-wise angle differences. In Figure \ref{fig:dist_upper} we show $P[\Sigma_{cos}(u_\perp,\Theta,N)]$ for randomly distributed $\Delta \ell_{j,k}$ for a variety of $N$, $u_\perp$, and $\Theta$ where the minimal $u_\perp$ is set by the maximal scale accessible by an interferometers primary beam, $\sim 1/\Theta$. We calculate these distributions from 10000 random realizations. We see that the distribution of $\Sigma_{cos}$  is independent of $N,\Theta$, and $u_\perp$ and has a mean of $\approx 0$ (which is the quantity that sets the amplitude of $\Sigma_{cross}$.   As long as sources are randomly distributed, we can expect LoS cross power spectra to suppress the cross terms sum to below the $10\%$ level at $\kpara \gtrsim 10^{-1} \Mpci$, regardless of the number of terms.

We may finally write. 
\begin{align} \label{eq:sum_forest_sourcesapprox}
P_f ( {\bf k} ) &\approx \displaystyle \sum_j   P_j(\kpara) \notag \\
 &= \frac{D_M^2 \Omega_{pix}^2}{\Omega_{cube}} \sum_j T_j^2 P_{\tau_{21}}^{LoS} \notag \\ 
&= \frac{D_M^2 \lambda^4}{4 k_B^2 \Omega_{cube}} \sum_j s_j^2 P_{\tau_{21}}^{LoS}
\end{align}
where $\lambda=\lambda_{21}(1+z)$ is the wavelength at the center of the data cube, $P_j$ is the absorption power spectrum for the $j^{th}$ source, $s_j$ and $T_j$ are the flux and temperatures of the $j^{th}$ source, $\Omega_{cube}$ is the solid angle subtended by the observed volume, and $P_{\tau_{21}}^{LoS}$ is the 1D LoS power spectrum.

 We may therefor consider the absorption power spectrum resulting from the forest as simply the sum of the absorption power spectra of each individual source in the background of the source cube. Since all quantities in this sum are positive, we see that the amplitude of the power spectrum increases linearly with the number of sources present behind an observed volume.  Because the power spectra for unresolved sources are constant in $k_\perp$, $P_f$ will have a structure that is nearly constant in $k_\perp$. 
 
Hence, for $k \gtrsim 10^{-1} \Mpci$, Equation (\ref{eq:sum_forest_sources}) simplifies to a sum of the auto power spectra along the LoS to each source. We finish by briefly commenting on the of the effect of clustering which we have ignored but we find (after comparing Equation (\ref{eq:pf_simple}) to our simulations) is still significant. Clustering will cause a disproportionate number of sources to reside in close proximity on the sky. The effect of this is two fold. First, the clustered sources will tend to be behind correlated optical depth columns so that the cross terms between such sources will be better described by auto power spectra. Second, the phases between such sources will be small so that they will not sum to zero. In addition, they will not introduce significant $\kperp$ structure except at the smallest perpendicular scales. Hence the cross terms introduced by clustered sources will closely resemble the $\kpara$ invariant auto terms and simply increase the overall amplitude of $P_f$. We treat this increase by introducing a multiplicative constant of order unity, $A_{cl}$, in Equation (\ref{eq:integral_fix}).

\section{A Comparison between two source models}\label{app:compare}
\begin{figure}
\includegraphics[width=.48\textwidth]{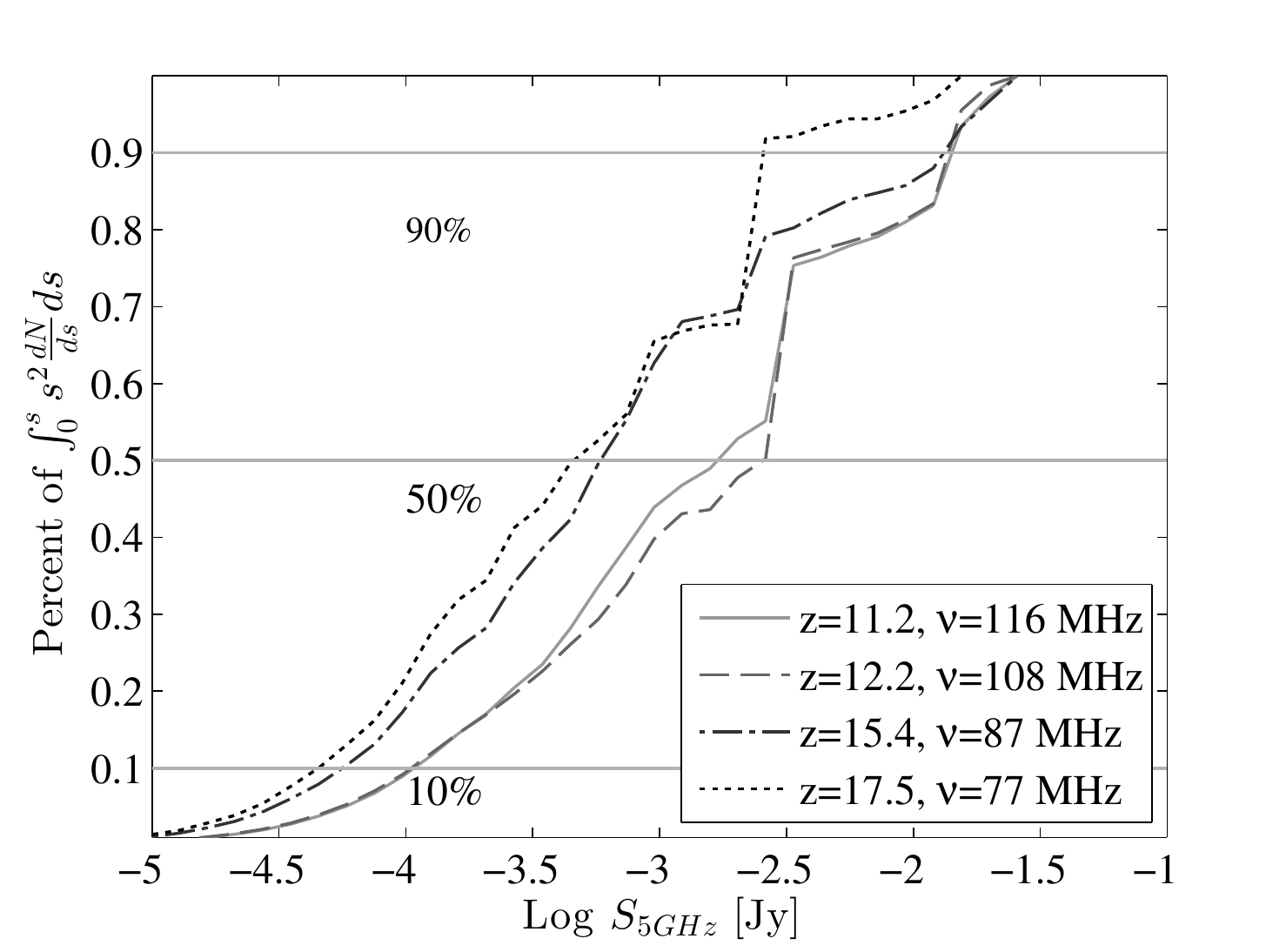}
~
\includegraphics[width=.48\textwidth]{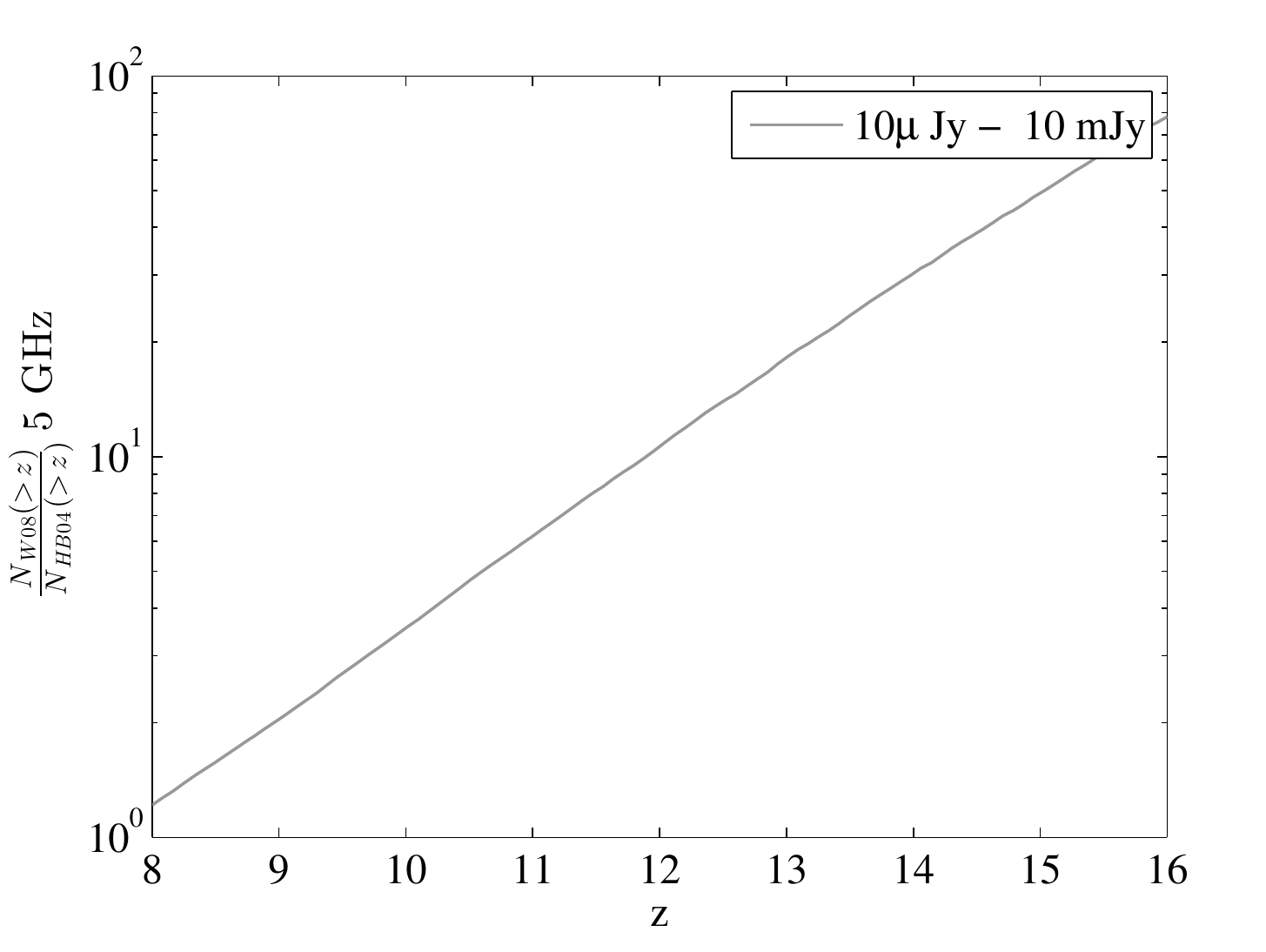}
\caption{Left: The percentage of the integrated luminosity function in Equation (\ref{eq:integral}) as a function of the source fluxes at $5 \text{GHz}$ for comparison to the catalogue of H04. We see that most contributions to the forest power spectrum come in between $S_{5 \GHz} = 10 \muJy$ and $S_{5 \GHz} = 10 \mJy$. Right: The ratio of the number of sources with redshift greater than $z$ between $S_{5\GHz} = 10 \muJy$ and $10 \mJy$ as predicted by the W08 and H04. The W08 simulation over predicts the counts in H04 by a factor ten at $z \gtrsim 12$ and nearly $80$ at $z \gtrsim 16$, emphasizing the importance of exploring this widely unconstrained parameter space in future work.}
\label{fig:flux_counts}
\end{figure}

In this paper, we choose to work with the semi-empirical source population in the simulation by \citet{2008MNRAS.388.1335W}. This choice was in part motivated by the lack of constraints at high redshift and the ease which which we could use data from the W08 simulation using its online interface. Another prediction in the literature for the high redshift radio luminosity function is made by \citet{2004ApJ...612..698H}. This model, like the one in W08, relies on a number of uncertain assumptions but is a more physically motivated bottom up approach which is derived from the cold dark matter power spectrum and assumptions about the black hole-halo mass relation and radio loud fraction. In this appendix, we attempt to understand how our choice of the Wilman source population compares to that in H04. To do this, we attempt to compare the source counts from W08 that contribute the most to $P_f$ to those of H04 who provide cumulative flux counts for $1-10 \GHz$ as a function of redshift. To compare the W08 sources, we compute the percentage of the radio luminosity function integral in Equation (\ref{eq:integral}) as a function of the extrapolated $S_{5 \GHz}$. On the left, in Figure \ref{fig:flux_counts}, a large fraction of $P_f$ is determined by W08 sources with $5 \GHz$ fluxes between $10 \muJy$ and $10 \mJy$. We show, in Figure \ref{fig:flux_counts}, the ratio of W08 and H04 source counts with $S_{5 \GHz}$ between $10 \muJy$ and $10 \mJy$. The H04 counts fall much faster with redshift than those of W08. At $z \sim 10-12$ the number of contributing sources is larger by a factor of $\approx 10$ and $\approx 80$ by $z \sim 16$.

This comparison is very approximate since different spectral indices are assumed in H04 and W08. However, we emphasize that the observability claims we make in this paper would not apply accurately to the H04 prediction. A more extensive exploration of parameter space will be necessary to determine what range of radio loud source populations may be constrained by the power spectrum technique. 

 Since $P_b$ is observed to be flat out to $k\approx 10 \Mpci$ while $P_f$ climbs, a more pessimistic source scenario has the effect of pushing the forest dominant region to higher $\kpara$ which does not preclude detection with a more powerful telescope such as the SKA.

\end{document}